\documentclass[reprint, superscriptaddress, amsmath, amssymb, aps, prb]{revtex4-2}

\UseRawInputEncoding

\usepackage{graphicx}
\usepackage{dcolumn}
\usepackage{bm}
\usepackage{xcolor}
\usepackage{hyperref}
\hypersetup{
    colorlinks=true,
    linkcolor=blue,
    filecolor=cyan,      
    urlcolor=cyan,
    citecolor=cyan,
  }
\usepackage{dsfont}

\usepackage{braket}
\newcommand{\rangll}{\rangle\hskip-0.5ex\rangle}

\newcommand{\ketbra}[2]{\ket {#1} \hskip -0.8ex \bra {#2}}
\newcommand{\kett}[1]{|#1\rangll}

\newcommand{\expct}[1]{\langle #1 \rangle}

\newcommand{\DAOE}{\text{DAOE}}

\DeclareMathOperator{\tr}{tr}

\begin{document}

\preprint{APS/123-QED}

\title{Open-system Spin Transport and Operator Weight Dissipation in Spin Chains}

\author{Yongchan Yoo}
\affiliation{
Department of Physics, Condensed Matter Theory Center, and Joint Quantum Institute,
University of Maryland, College Park, Maryland 20742, USA}

\author{Christopher David White}
\affiliation{Joint Center for Quantum Information and Computer Science, University of Maryland, College Park, Md, 20742}
\affiliation{Condensed Matter Theory Center, University of Maryland, College Park, Md, 20742}

\author{Brian Swingle}
\affiliation{Department of Physics, Brandeis University, Waltham, Massachusetts, 02453}

\date{\today}

\begin{abstract}
We use non-equilibrium steady states to study the effect of dissipation-assisted operator evolution (DAOE) on the scaling behavior of transport in one-dimensional spin chains.
We consider three models in the XXZ family: the XXZ model with staggered anisotropy, which is chaotic;
XXZ model with no external field and tunable interaction, which is Bethe ansatz integrable and (in the zero interaction limit) free fermion integrable;
and the disordered XY model, which is free-fermion integrable and Anderson localized.
We find evidence that DAOE's effect on transport is controlled by its effect on the system's conserved quantities.
To the extent that DAOE preserves those symmetries, it preserves the scaling of the system's transport properties;
to the extent it breaks those conserved quantities, it pushes the system towards diffusive scaling of transport.

\end{abstract}

\maketitle

\section{Introduction}

Quantum out-of-equilibrium dynamics is at the heart of various areas of physics from condensed matter to high energy physics and even quantum information science. The dynamics of conserved quantities is particularly interesting within this broad non-equilibrium setting. In the solid-state context, measurements of transport of conserved quantities like energy and charge provide a useful window into the underlying dynamics of these complex systems. In particular the \textit{scaling behavior} of a system's transport properties---whether it is diffusive, subdiffusive, or superdiffusive, as well as details like the nature of the scaling function---is intimately connected with the strength of the system's interactions \cite{DGV+22}, the presence of kinetic constraints and higher-form symmetries \cite{SWV+21, ILN21, GLN20, FSD+20, MKH20, IVN19, GGR+22, SVG23, RP22, SLR+21, McG22}, and its integrable or chaotic nature. Recent experimental developments in various platforms including cold atom systems~\cite{BDZ08, BDN12, JAD+20, GTH+21, SWB+22, WRY+22}, quantum magnets \cite{SSD+21}, superconducting quantum circuits~\cite{BGG+21}, and heavy-ion collisions~\cite{BRS18} are also shedding light on the subject. Along with those experimental results, new theoretical approaches have been developed to tackle the major challenge of calculating and interpreting the observed transport phenomena. Due to the breadth of the subject, theoretical developments include a range of approaches from general frameworks to techniques for specific situations (reviews include~\cite{CCG+11, PSS+11, ISG12, GBL13, EFG15, CEM16, GE16, AAB+19, BHK+21}).

These new approaches are especially important for strongly interacting systems where the physical interpretation of transport phenomena is not well understood. Numerical approaches are indispensable since there is often no simple analytical technique available. Tensor network algorithms, especially matrix product state methods, can access transport physics close to the thermodynamic limit~\cite{Whi92, Sch11, PKS+19}. For other commonly considered problems (e.g. ground states of gapped local Hamiltonians and short-time evolution), matrix product state methods are reliable because the states in question have low entanglement. For short-time evolution in particular, TEBD \cite{Vid03, Vid04} constitutes a controlled approximation. But matrix product state methods become expensive for systems with slow dynamics (e.g., subdiffusive transport~~\cite{ZSV16, STS+20}) or high amounts of entanglement. Some alternate techniques have been suggested~\cite{LPB+17, KLR18, HLB+18, WZM+18, KEL19, YMW+20, RVP20, KHB21, LSA22, WP18, WPS18, ZRS19}. Many of those methods modify the dynamics to a non-unitary time evolution not unlike a Lindblad dynamics. By doing so, they cut off (notionally) less relevant parts of the dynamics while preserving the essential transport physics. From a tensor network perspective, one important outcome of the modification is to reduce the amount of entanglement while preserving the physics of interest. Developing a principled theory of when and why these methods work is an active line of research \cite{VPR21, Whi21, NRV+22}.

One of these new tensor network methods, \textit{dissipation assisted operator evolution} (DAOE)~\cite{RVP20} employs an artificial dissipation based on \textit{operator weight} to overcome the entanglement barrier in unitary simulations. Here operator weight refers to the number of non-identity single-qubit operators contained in a many-body operator; suppressing high-weight operators---that is, suppressing many-point correlations---suppresses many-body entanglement. Because the conserved quantities and their currents are local operators, the artificial dissipation does not directly modify those quantities or currents. In a chaotic system, the expectation values of conserved quantities and currents determine the state of the system, so one expects the artificial dissipation not to substantially modify the system's state or dynamics. Moreover, because DAOE directly manipulates the operator weight distribution, it is possible to study the influence of operator growth~\cite{NVH18, RPV18, KVH18, GHK+18, CDC18, PCA+19} on transport physics. Fig.~\ref{fig:schem} gives a schematic of DAOE as implemented with matrix product operators.

We investigate the effect of operator weight dissipation on the scaling behavior of spin transport in one-dimensional lattice models by combining two sources of non-unitarity: DAOE and boundary-driven open system dynamics. The physical quantity of interest is the scaling exponent relating the average spin current to the system size. Diffusive transport gives one characteristic value of the exponent, and the exponent allows us to characterize the transport away from the diffusive case.

First, as a benchmark, we apply the method to the anisotropic XXZ model with a staggered field, which is chaotic and possesses normal diffusive relaxation of the spin current. We find that for any operator dissipation parameters, the normal diffusive transport is maintained. Next, we study the clean XXZ model in three different regimes. In the weak interaction regime, the modified transport shows superdiffusive transport up to the system size we calculated ($N \sim 256$), whereas the unitary limit is believed to exhibit ballistic transport~\cite{JZ11, LHM+11, SVO18, DM20}. At the isotropic point ($\Delta = 1$) where the non-dissipative transport exhibits a superdiffusive relaxation, the transport under DAOE is still superdiffusive but the scaling exponents vary depending on the operator cut-off length. In the strong interaction regime ($\Delta > 1$), the unitary system's diffusive transport is retained for all operator cut-off lengths up to the largest system size. Lastly, we treat the disordered XY model. There we observe behavior consistent with coherent transport on length scales given by the DAOE cut-off length and diffusive transport on longer length scales; we explain this in terms of DAOE's effect on Anderson orbitals.

Taken together, these results point to the following conclusions. First, as a technical point, DAOE can be usefully combined with open system dynamics. This introduces a need to extrapolate to the physical limit, but the NESS is generally easier to obtain and more stable in the presence of artificial dissipation. Second, DAOE tends to push the dynamics towards diffusive transport, all other things being equal. It maintains diffusivity for generic chaotic models and typically breaks integrability in non-chaotic models. Third, how well DAOE captures the underlying unitary dynamics depends sensitively on the number and nature of the symmetries it preserves. We elaborate on these points in the discussion.

The rest of the paper is structured as follows. In Section II we introduce the model and the framework for analyzing spin currents. Next, in Section III we describe the methods combining DAOE with open system dynamics. In Section IV we present our main results which include various one-dimensional spin models and crossovers between different transport types. Lastly, we discuss the results and possible future directions in Section V.

\section{Model and quantities of interest}

\subsection{Model}

We study spin transport in three variations of an XXZ spin chain. The general form of the Hamiltonian is
\begin{subequations}
\label{eq:model}
\begin{align}
    H &= \sum_{i = 1}^{N - 1} H_{i,i+1}, \\
    H_{i,i+1} &=  \sigma_i^x \sigma_{i + 1}^x + \sigma_i^y \sigma_{i + 1}^y + \Delta \sigma_i^z \sigma_{i + 1}^z\notag\\
    &\qquad+ \frac{1}{2} (h_i \sigma_i^z + h_{i + 1} \sigma_{i + 1}^z), 
\end{align}
\end{subequations}
where $\sigma^\alpha_i$'s are Pauli matrices, $\Delta$ controls the anisotropy, and $h_i$ is the magnitude of $z$-directed field at site $i$.

The model \eqref{eq:model} displays a rich variety of spin-transport behaviors. At $\Delta = 0$ it is free-fermion integrable, so it displays ballistic transport if the $h_j$ are uniform and Anderson localization if the $h_j$ are random. It can also exhibit a transition between the two behaviors if the $h_j$ are appropriately quasiperiodic \cite{AA80}.

For $\Delta \ne 0$ and $h_j = 0$ uniform, the model is Bethe-ansatz integrable. At half filling it is ballistic for $\Delta \le 1$ and diffusive for $\Delta > 1$ \cite{GV19, DBD19, LZP17, MKP17, ID17, BVK+17, KMH14, DCD17, IP13, PPS+14, Pro11, ZNP97, PZ16, CDV18, Zot99, HPZ11, IDM+18, KKM14, MKP17-1, SGB14, KPS15, AKT06, Ste12, DS98, DS05, PZ13}. (\cite{DBD19} Sec. 6 has a useful, concise summary of this literature.) At the isotropic point $\Delta = 1, h = 0$ the model is $SU(2)$ symmetric; this symmetry appears to protect the superdiffusive behavior, which remains even for large $SU(2)$-symmetric perturbations \cite{DGV+21}.

For $\Delta \ne 0$ and $h_i$ random, the spin transport is not well understood. For small to moderate disorder, the model appears to display anomalous diffusion \cite{BCR15, AGK+15, TS15, LLA16, ZSV16, PH17, MZV+19, LKL20}.
This anomalous diffusion may be due to Griffiths rare region effects~\cite{Gri69, VHA15, PVP15, GAD+16, AAD+17}, but other possible scenarios include irregular scaling of the matrix elements~\cite{LB16}, multifractality of eigenstates~\cite{LKB20}, and a non-Griffiths phenomenological theory of the resistance distribution~\cite{STS+20}. The system may undergo a many-body localization (MBL)~\cite{BAA06, ZPP08, PH10} transition at $h \approx 7.6$, but recent work has cast doubt on the location and indeed existence of the transition~\cite{SBP+20, ABD+21, SP21, GZ22, GZ22-1, MCK+22, SZ22, CC22}. Detailed reviews are available for the MBL phases~\cite{NH15, AL18, AAB+19}.

In this paper, we consider three parameter regimes: one chaotic, one Bethe ansatz integrable, and one Anderson localized. DAOE \cite{RVP20} was designed to compute transport coefficients in the first regime, for chaotic one-dimensional quantum systems. To test the method in this case we consider the anisotropic XXZ model with anisotropy $\Delta = 0.5$ and staggered field $h_{2i} = -0.5$ and $h_{2i + 1} = 0$. With these parameters, the model is non-integrable and shows diffusive transport~\cite{PZ09}.

We then pick two cases to study anomalous transport: the zero-field XXZ model and the disordered XX model. The XXZ model at zero field disorder $h = 0$ exhibits various transport types as the anisotropy $\Delta$ increases from zero. For weak anisotropy $\Delta < 1$, one finds ballistic transport; in the opposite regime $\Delta > 1$, the model exhibits normal diffusive transport. The critical point is at the isotropic point, $\Delta = 1$, where superdiffusive but subballistic transport occurs. The disordered XY model ($\Delta = 0$, $h \neq 0$) also displays an interacting crossover in its transport. The clean limit has ballistic transport thanks to a dual free fermion description, whereas any non-zero disorder brings Anderson localization in the thermodynamic limit~\cite{And58}. But the model always possesses an extensive number of conserved quantities, and the physical size of these conserved quantities in the spin language varies with the disorder strength.

\subsection{Spin Current Analysis}\label{ss:spin-current-analysis}

\begin{figure}[t]
\includegraphics[width=0.9\linewidth]{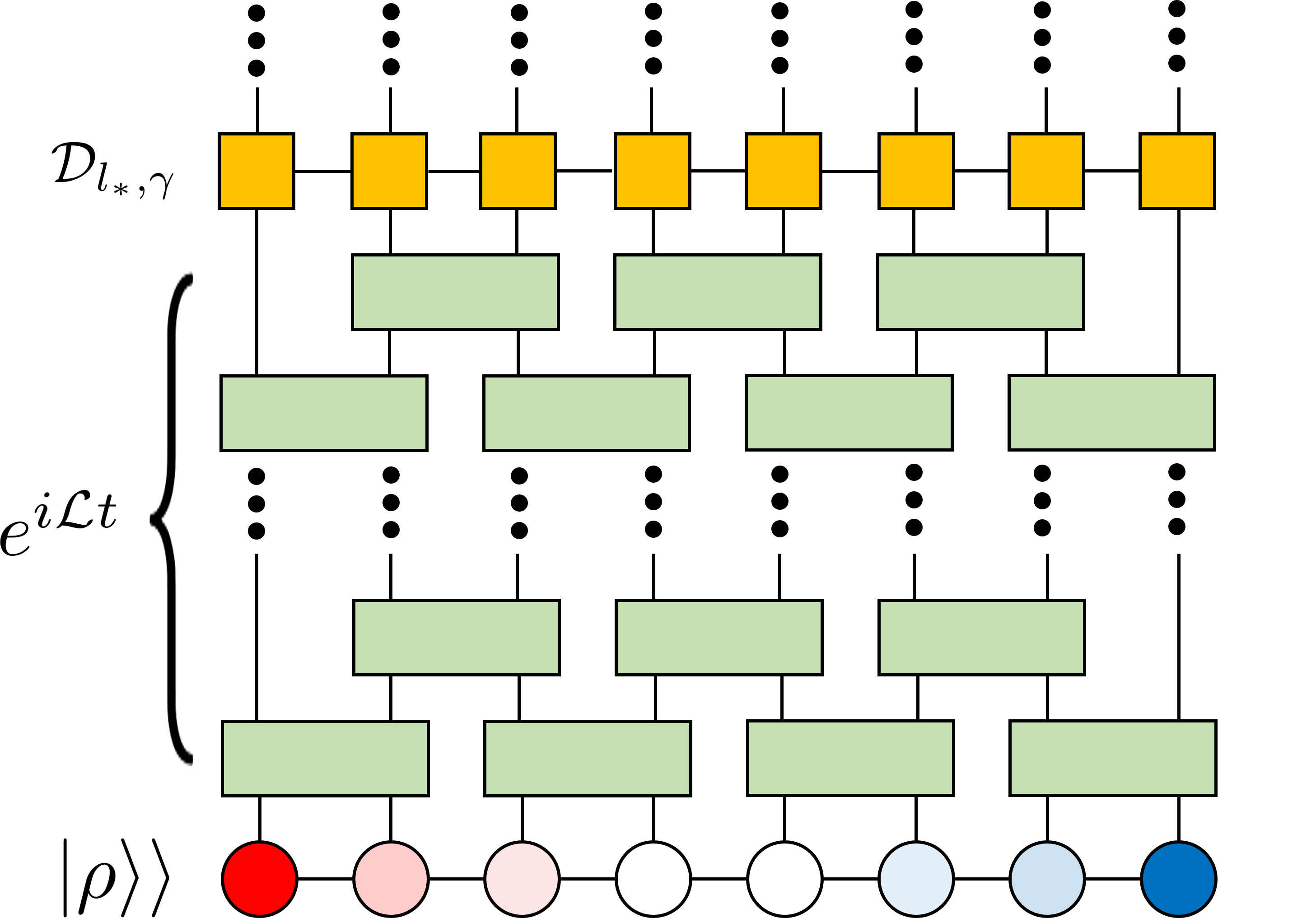}
\centering
\caption{Schematics of the combined method of the boundary driven open quantum system and DAOE. It describes one period of the artificial dissipation superoperator application. The gradation of $\lvert \rho \rangle \rangle$ from red to blue is for the spin imbalance by the Markovian spin bath setting.}
\label{fig:schem}
\end{figure}

Suppose a system has a conserved quantity $Q = \sum_i Q_i$, $Q_i$ local. The corresponding local current $J_i$ is derived from the continuity equation and the Heisenberg equations of motion:
\begin{equation}
    \frac{\partial Q_i}{\partial t} = -i \left[ Q_i, H \right] = - (J_i - J_{i + 1}).
\end{equation}
The model \eqref{eq:model} has a conserved quantity $Q^z = \sum_i \sigma_i^z$, the total $z$-spin; the current is $J_i = \sigma_i^x \sigma_{i+1}^y - \sigma_i^y \sigma_{i+1}^x$.

For systems exhibiting diffusive transport, the discrete Fourier's law $\langle J_i \rangle = -D( \langle Q_{i+1} \rangle - \langle Q_i\rangle)$ relates the current and the corresponding charge density. Here $D$ is the diffusion constant in lattice units. If such a diffusive system is subject to a bias $\langle Q_L \rangle - \langle Q_R\rangle$, where $\langle Q_{L,R}\rangle$ denote fixed values of the spin density at the left and right ends of the sample, the current through the sample scales as
\begin{equation}
    \langle J\rangle = -D \frac{\langle Q_L \rangle - \langle Q_R \rangle}{N},
\end{equation}
where $N$ is the length of the system.

More generally, if the system exhibits anomalous transport, the above relation is modified by introducing a scaling exponent $\chi$,
\begin{equation} \label{eq:JvsN}
    \langle J\rangle = -D_\chi \frac{\langle Q_L \rangle - \langle Q_R \rangle}{N^\chi}
\end{equation}
We assume that $\chi$ in Eq.~\eqref{eq:JvsN} is the only scaling exponent that characterizes the transport. Other than the normal diffusive transport ($\chi = 1$), possible types of anomalous transport are (i) ballistic transport ($\chi = 0$), (ii) superdiffusive transport $(0 < \chi < 1)$, and (iii) subdiffusive transport $(\chi > 1)$. In a localized state, this power-law ansatz does not provide a good description of the spin transport. We can heuristically understand localized systems as having $\chi \to \infty$.

\section{Methods}

We extract quantum transport properties with dissipation assisted operator evolution (DAOE) simulations
of non-equilibrium steady states (NESS). Each method takes advantage of non-unitary evolution to make simulating a system's dynamics tractable. We find that combining them gives new insights into both the systems' physics and the effect of DAOE on that physics. In this section we describe the two methods.

\subsection{Master Equation and NESS}

In a NESS experiment on a spin chain we attach leads with slightly different chemical potentials to the two ends of the system. The leads thermalize the system, so in the long-time limit its state should have an efficient MPO representation\cite{Has06, SVC+15}. But because the leads' chemical potentials differ, they induce a small spin current; how this current scales with system size characterizes the model's transport properties (cf Sec.~\ref{ss:spin-current-analysis}).

Formally the NESS is the fixed point solution $d \rho_\infty / dt = 0$ of the Gorini-Kossakowski-Lindblad-Sudarshan (GKLS) master equation~\cite{Lin76, GK76}
\begin{equation} \label{Lind}
    \frac{d \rho}{dt} = \mathcal{L} (\rho) \equiv i [\rho, H] + \sum_{\nu} \left [ L_{\nu} \rho L_{\nu}^{\dagger} - \frac{1}{2} \left \{ L_{\nu}^{\dagger} L_{\nu}, \rho \right \} \right].
\end{equation}
The NESS is generated by full Hamiltonian $H$ and Lindblad operators $L_{\nu}$, which model the leads. Explicitly, the Lindblad operators are 
\begin{align}
\begin{split}
    L_{1, \pm} &= \sqrt{1 \pm \mu}\; \sigma_{1}^{\pm}\\
    L_{N, \pm} &= \sqrt{1 \mp \mu}\; \sigma_{N}^{\pm}
\end{split}
\end{align}
where $\sigma^{\pm} = \frac{1}{2}(\sigma^{x} \pm i \sigma^{y})$.

Under the right conditions, the GKLS equation has exactly one steady-state solution~\cite{Nig19}, but even when this is the case, there may still be many slowly-decaying almost steady states, especially when the jump operators only affect the edges of the sample. We expect to have a unique NESS $\rho_\infty$ which is accessible in the long-time limit $t \rightarrow \infty$, but the presence of slow modes means we must be careful about convergence in time.

\subsection{Artificial Dissipation Superoperator}

Dissipation assisted operator evolution (DAOE) \cite{RVP20} is a tensor network-based algorithm that reduces the weight of operators longer than a given cut-off length
\footnote{
    NB in this context \text{length} means number of nontrivial Pauli operators. This is in contrast to \text{diameter}, or distance between leftmost and rightmost nontrivial Pauli operator, which \cite{Whi21} and (implicitly) \cite{VPR21} argue is the relevant quantity in spatially local systems.
    }.
These long operators are responsible for the entanglement growth that makes MPS simulations infeasible. By gently reducing them, DAOE makes long-time simulations possible.

DAOE is implemented by an artificial dissipation superoperator acting on the operator Hilbert space. The operator Hilbert space of our $N$-site system is spanned by a basis of $4^N$ Pauli strings. Each element (Pauli string) $\mathcal{S}$ in the basis is represented by the tensor product of single-site Pauli matrices $\sigma^0, \sigma^x, \sigma^y, \sigma^z$. The \textit{length} $\ell_\mathcal{S}$ of a string $\mathcal{S}$ is the number of non-trivial Pauli matrices in $\mathcal{S}$. In this notation the artificial dissipation superoperator is
\begin{equation} \label{DAO}
    \mathcal{D}_{\ell_\ast, \gamma} \left[ \mathcal{S} \right] = 
        \begin{cases}
        \mathcal{S} &\text{if }\ell_\mathcal{S} \leq \ell_\ast\\
        e^{-\gamma \left( \ell_\mathcal{S} - \ell_\ast \right)} \mathcal{S} &\text{if }\ell_\mathcal{S} > \ell_\ast.
        \end{cases}
\end{equation}
Periodically applying this superoperator generates a non-unitary quantum evolution that can be heuristically understood as a global `bath'. Just as a bath---consider in particular the depolarizing channel---reduces the expectation value of a string of $\ell$ nontrivial Pauli operators by an amount $\propto \ell$, the DAOE superoperator \eqref{DAO} reduces the expectation value of a string of $\ell$ nontrivial Pauli operators by an amount $\propto \max(\ell - \ell_*, 0)$. 

DAOE as presented in \cite{RVP20} uses the above artificial operator dissipation to modify the Heisenberg picture dynamics of observables. For example, starting from an initial state $\rho_0$ with some spatially varying profile for the average spin density $\tr( S^z_r \rho_0)$, the spin diffusivity can be extracted from the time-dependent spin profile $\tr(S^z_r(t) \rho_0)$ where $S^z_r(t)$ is the Heisenberg evolution of $S^z$ at site $r$. DAOE is then used to modify the dynamics of $S^z_r(t)$ to render it more tractable to an entanglement-constrained tensor network simulation, with the true physics obtained from an extrapolation in $\gamma$.

However, more than just modifying the particular dynamics with the introduction of $\gamma$, DAOE significantly alters the basic rules of quantum evolution. This is most easily seen in the Schrodinger picture formulation, where the fact that DAOE reduces expectation values of long operators without reducing expectation values of short operators means it can break positivity of density matrices. Consider applying $D_{\ell_* = 1, \gamma}$ to the density matrix of the two-site state $\ket{\uparrow\uparrow}$: it becomes
\begin{align}\label{eq:daoe-positivity-example}
\begin{split}
&\frac 1 4 \mathcal D_{\ell_*=1, \gamma}\Big[I + \sigma_1^z + \sigma_2^z + \sigma^1_z \sigma^2_z\Big] \\
&\qquad=I + \sigma_1^z + \sigma_2^z + e^{-\gamma}\sigma^1_z \sigma^2_z
\end{split}
\end{align}
which has one negative eigenvalue $\frac 1 4 (e^{-\gamma}-1)$.

Crucially, the Heisenberg and Schrodinger pictures remain equivalent even in the presence of DAOE's artificial operator dissipation
\footnote{
    Not all schemes have this property. The Liouvillian graph scheme of \cite{Whi21}---like DAOE---gives a linear effective evolution, so Heisenberg and Schrodinger evolutions are---as in DAOE---identical. But DMT \cite{WZM+18} is strongly nonlinear, because it uses the SVD of a correlation matrix to determine what correlations to discard. (This nonlinearity is likely responsible for DMT's success in treating nearly free fermion\cite{YMW+20} and integrable KPZ\cite{WRY+22, YMK+22} transport.) The method of Kvorning, Herviou, and Bardarson \cite{KHB21} is also nonlinear, because it approximates long-range correlation functions by products of local expectation values.
    }.
In the Heisenberg picture, the time evolution of some operator $A$ by a Lindbladian $\mathcal L$ becomes
\begin{align}
    A(t) = \left[\mathcal D_{\ell_*, \gamma} e^{-i\mathcal L \tau} \right]^{t/\tau}A(0)\;.
\end{align}
But the DAOE superoperator, like the Lindblad time evolution superoperator $e^{i\mathcal L\tau}$, is linear; indeed $\mathcal D_{\ell_*,\gamma}$ is Hermitian under the trace inner product $\langle A,B\rangle = \tr A^\dagger B$. So the operator expectation value $\langle A(t)\rangle$ obeys
\begin{align}
    \tr A(t)\rho(0) = \tr A \rho(t)
\end{align}
where $A$ is the Schr\"odinger picture operator and
\begin{align}
    \rho(t) = \left[\mathcal D_{\ell_*, \gamma}\rho(0) e^{i\mathcal L \tau} \right]^{t/\tau} \;.
\end{align}
Hence, the Schr\"odinger and Heisenberg pictures give identical dynamics, so we can speak of DAOE ``failing to preserve positivity''. Moreover we can use NESS simulations to examine how DAOE changes a system's dynamics, with full confidence that the results apply to Heisenberg-picture experiments like those of \cite{RVP20}.

Given the significant modifications that DAOE makes to the quantum dynamics, it is important to understand when and why DAOE gives a good approximation of the transport coefficients. Ref.~\cite{VPR21} analyses the ``operator backflow'' process, in which information contained in large-diameter, non-local operators flows into the subspace of short operators as the system evolves. For chaotic models, combinatoric scattering amplitude arguments and numerical experiments confirm the exponential suppression of the backflow process contribution to correlation functions between local operators. Consequently, the estimated error of DAOE-produced transport coefficients is also exponentially small in such systems.

But this backflow analysis will not go through for integrable systems. In the language of \cite{Whi21}, the backflow analysis assumes that the dynamics of long operators is chaotic. But in the integrable system the tower of local conserved quantities will strongly constrain that dynamics---it cannot be treated as chaotic. Additionally, the interplay of DAOE with more complex non-Abelian symmetries and with integrability has not yet been studied. Some of our results below address these open problems.

\subsection{Tensor Network Implementation}

\begin{figure*}
\centering
\includegraphics[width=\linewidth]{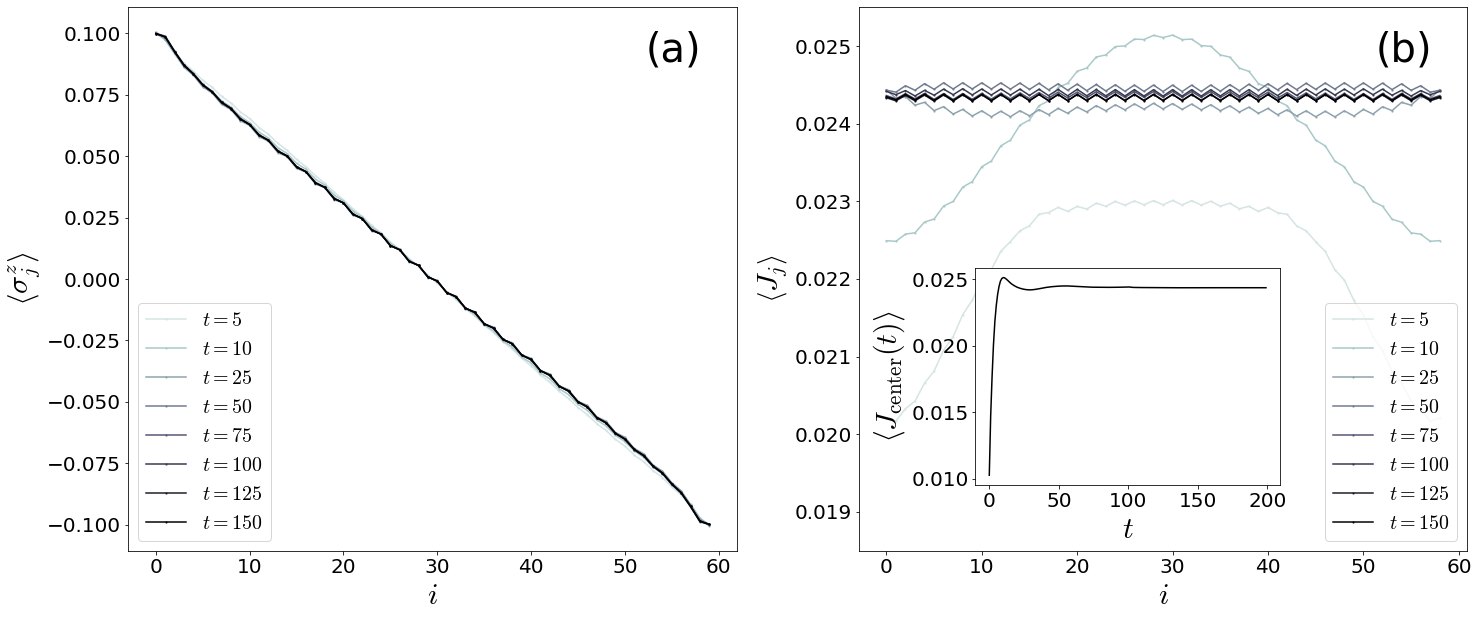}
\centering
\caption{(a) Spin profile and (b) spin current profile extracted from the modified NESS of the chaotic anisotropic XXZ model with $(L, \ell_*, \gamma) = (60, 3, 0.4)$ at different times. The inset in (b) shows the time evolution of the spin current at the center of the chain.}
\label{fig:SnJprof}
\end{figure*}

Both NESS and DAOE can be efficiently realized in the language of tensor networks. In this formalism, a vector in the operator Hilbert space directly expresses the corresponding density matrix; we call such a vector a \textit{superket state} $\kett \rho$. Physical operators can act on $\rho$ in two ways, ``bra-side'' and ``ket-side''---formally these are the left and right regular representations of $GL\Big[(\mathds C^2)^{\otimes n}\Big]$. Two different physical operators $X,Y$ can act on $\rho$ by $\kett{X \rho Y} = Y^T \otimes X \kett \rho$---formally, this is a representation of $\Big(GL\Big[(\mathds C^2)^{\otimes n}\Big]\Big)^{\otimes 2}$. In this representation the Lindbladian operator in Eq. \eqref{Lind} is
\begin{align} \label{Lindvec}
\begin{split}
    \mathcal{L} &= -i \left(I \otimes H - H^T \otimes I \right) \\
    &\qquad +  \sum_{\nu} \left ( L_{\nu}^{\ast} \otimes L_{\nu} - \frac{1}{2} \left ( I \otimes L_{\nu}^{\dagger} L_{\nu} + L_{\nu}^T L_{\nu}^* \otimes I \right) \right).
\end{split}
\end{align}
Since the operator Hilbert space has a tensor product structure we can implement superket-superoperator calculations with standard MPS equipment. The time evolution of the superket is performed by the standard time-evolving block decimation (TEBD) algorithm \cite{Vid03, Vid04} with second-order Suzuki-Trotter decomposition of the time evolution operator $e^{\mathcal{L} t}$ \cite{Tro59, Suz76, PKS+19}. We use the Trotter time step $\delta t = 0.1$ for our numerical simulations and check convergence in the Trotter step in App.~\ref{app:conv}.

We interleave the TEBD Lindbladian evolution with periodic application of the DAOE superoperator (cf Fig.~\ref{fig:schem}). We take the DAOE period $\tau = 1$. For small $\gamma$, the ratio $\gamma/\tau$ controls the effective dissipative dynamics; in principle, one would like to take $\tau$ small. The DAOE superoperator $\mathcal D_{\ell_*,\gamma}$ has an exact MPO representation of bond dimension $\ell_*$, so applying it to a matrix product density operator is not infeasible. But $\mathcal D_{\ell_*,\gamma}$ does not conveniently commute with the terms in the Lindbladian \eqref{Lindvec}, so we cannot fold it into Suzuki-Trotter decomposition leading to TEBD, and we treat it separately. At time $t \in \left[N\tau, (N + 1)\tau \right]$, the initial superket $\kett{\rho (0)}$ is evolved into
\begin{equation}
    \kett{\rho (t)} = e^{\mathcal{L} (t - N \tau)} \left( \mathcal{D}_{\ell_\ast, \gamma} e^{\mathcal{L} \tau} \right)^N \kett{\rho(0)}
\end{equation}
We call the resulting state the modified NESS to stress the dissipation of long operator contributions. 

For some models---especially those with subdiffusive or localized transport---the state can take a very long time to converge to the (modified) NESS. To alleviate this problem, we choose the initial superket as a linearly interpolated state between the two baths at the end of the system. Explicitly, the initial superket is chosen as $\lvert \rho(0) \rangle \rangle \propto \exp(-\sum_i \mu_i \sigma_i^z)$. Since this spin profile is similar to the steady-state solution of Fourier's law we expect fast convergence to the modified NESS.

After obtaining the modified NESS, we directly calculate the expectation value of a local operator $\mathcal{O}$ by taking the operator trace $\langle \mathcal O \rangle = \text{tr}(\mathcal O \lvert \rho_\infty \rangle \rangle) / \text{tr}(\lvert \rho_\infty \rangle \rangle)$. There are small fluctuations in site $i$ and time $t$ of the current $J_i(t)$ even if we closely approach the modified NESS due to the limitation of our numerical methods. To avoid this issue, we average over all sites and a small time window to estimate the average current $J$ which is supposed to be independent of $i$ and $t$.

\subsection{Convergence}

We check the convergence of the normalized, spatially-averaged current expectation value in the following three categories.
\begin{itemize}
    \item Time: We declare that the NESS has been approximately reached when the relative change in the current is less than $10^{-4}$ per unit of time ($J^{-1}$).
    \item Bond dimension: the current varies $< 2\%$ between the bond dimension shown and a bond dimension smaller by a factor of $1/2$ or $3/4$ (depending on model and parameters).
    \item Trotter step size: for the clean models the current varies $< 1\%$ between the Trotter step shown and a smaller Trotter step size $\delta t = 0.025$ ($1/4$ the value used in plots shown).
    We allow larger tolerances for the disordered $XY$ model: there the worst case is $8\%$ at the strongest disorder ($h > 3.0$).
\end{itemize}

The total error budget is thus less than $5\%$ for the clean models and $10\%$ for the worst case of the disordered model. In general, convergence is affected by the DAOE parameters: it becomes worse as the dissipation strength $\gamma$ decreases and the cut-off length $\ell_\ast$ increases. For a fixed set of TEBD and DAOE parameters, we also find that it becomes harder to obtain a reliable NESS for larger anisotropy $\Delta$ (disorder $h$) in the clean XXZ (disordered XY) model. Details are presented in App.~\ref{app:conv}.

\section{Results}

\begin{figure}[t]
\includegraphics[width=\linewidth]{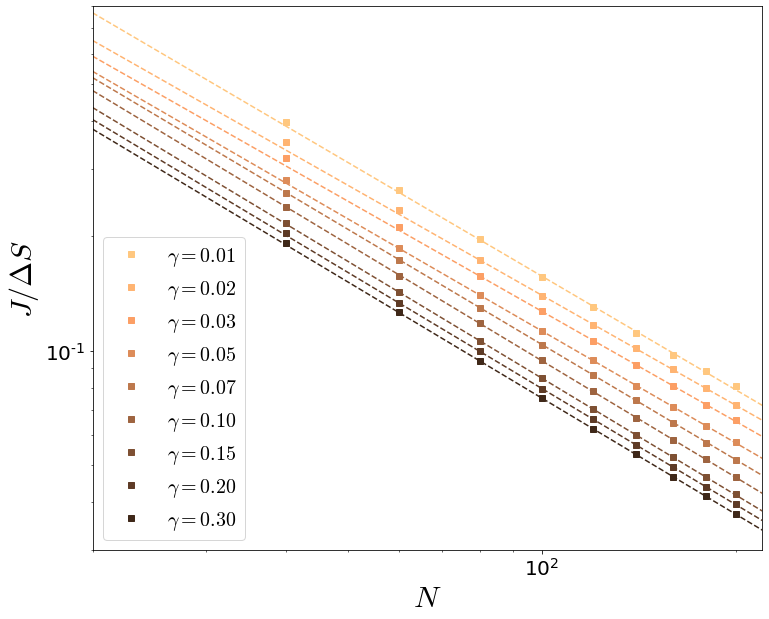}
\includegraphics[width=\linewidth]{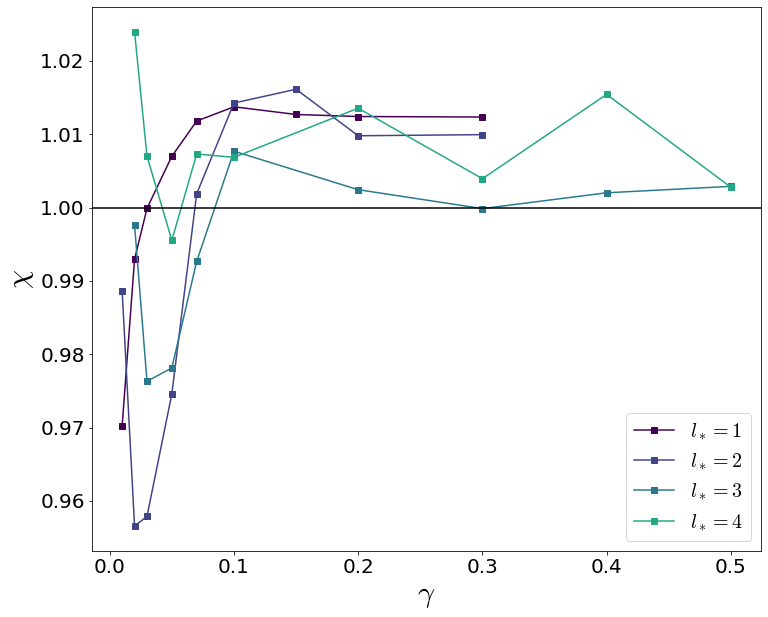}
\centering
\caption{
\textbf{Top}:
Scaled average spin current with DAOE cut-off length $\ell_\ast = 2$ for the chaotic anisotropic XXZ model with a staggered field as a function of system size $N$. The dashed lines are the best power law that fits the data with system size $N \ge 100$. Although the data is not shown, similar results hold for all other cut-off lengths.
\textbf{Bottom}: 
Scaling exponent extracted from the fit. The black horizontal solid line represents $\chi$ corresponding to the (normal) diffusion. The model parameters are in the main text.}
\label{fig:SJvsN}
\end{figure}

\subsection{Chaotic Anisotropic XXZ Model}

First, we study spin transport in the anisotropic XXZ model with a staggered field. We take the anisotropy $\Delta = 0.5$ and the staggered field $h_{2i} = -0.5, h_{2i+1} = 0$; the staggered field breaks integrability. We impose a chemical potential difference 
\[ \mu = 0.1 \]
and work at bond dimension 32. We check that the system has in fact converged to the NESS by comparing the variance of the spin current across sites to its average.

Fig.~\ref{fig:SnJprof} shows the spin $\expct{\sigma^z_j}$ and current $\expct{J_j}$ as a function of position in the NESS of this model. It displays roughly the expected linear profile. The zigzag pattern in $\expct{\sigma^z_j}$ comes about because of the staggered field. Additionally,  the interplay of the Hamiltonian bond term with the Lindblad operators causes $\expct{\sigma^z_j}$ to depart from the linear profile near the boundaries. We, therefore, drop 5 sites at the left and right end of the chain, so
\begin{align}
    \Delta S = \expct{\sigma^z_6} - \expct{\sigma^z_{N - 5}}
\end{align}
Likewise in fits to \eqref{eq:JvsN} we take the length to be $N - 10$.

Fig.~\ref{fig:SJvsN} top shows $J/\Delta S$ as a function of system size across dissipation strengths $\gamma$, all for cutoff $\ell_* = 2$, together with power-law fits. The fits are solely to system sizes $N \ge 100$. Fig.~\ref{fig:SJvsN} bottom shows the power resulting from the fit, across artificial dissipation strengths $\gamma$ and cutoffs $\ell_*$. The powers are all close to one. They deviate more for smaller $\gamma$ because DAOE reduces large correlation functions less quickly at smaller $\gamma$ so the simulations are more computationally demanding. The simulations do converge within $3\%$ for NESS expectation values at bond dimension 32, but those criteria leave room for the small deviations from the diffusive exponent $\chi = 1$ that we see in Fig.~\ref{fig:SJvsN} bottom.

To see why small-$\gamma$ simulations are more difficult---and why large-$\gamma$ simulations modify transport---we can make a rough model for the dynamics of the operator length distribution. The Hamiltonian increases the length of a Pauli string at a rate $\approx 1$ (in our units). DAOE, on the other hand, decreases weight on a Pauli string of length $\ell$ at a rate $\approx \gamma (\ell - \ell_*)$. These effects balance at a characteristic DAOE lengthscale $\ell_\DAOE \sim \ell_* + 1/\gamma$.

But even in the unitary NESS, the boundary Lindblad operators (and the resulting spread of entanglement) limit operator growth; this is one way to see why unitary NESS simulations are feasible. Write $\xi$ for the characteristic operator length of the unitary NESS. $\xi$ is not quite the correlation length: long operators contribute substantially to $\xi$ even if they have a small weight, because there are many of them.

When $\gamma \ll (\xi - \ell_*)^{-1}$, DAOE does not substantially change the operator weight dynamics, because the spread of bath entanglement keeps most operators shorter than the characteristic DAOE length $\ell_\DAOE$. The simulations are then approximately as hard as unitary NESS simulations. When $\gamma \gg (\xi - \ell_*)^{-1}$, by contrast, DAOE substantially reduces the characteristic operator length and the simulations become easier than unitary NESS simulations.

\begin{figure}[t]
\includegraphics[width=\linewidth]{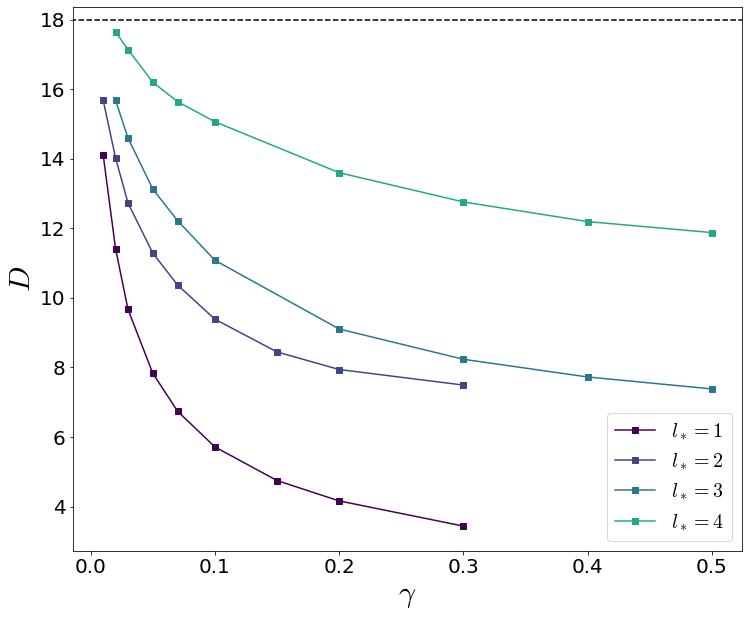}
\centering
\caption{Diffusion constants of the chaotic anisotropic XXZ model for various DAOE parameters, from fits with fixed $\chi = 1$. The black horizontal dashed line represents $D$ at the unitary limit ($\gamma \rightarrow 0$) from Ref. ~\cite{PZ09}.}
\label{fig:SChi}
\end{figure}

In Fig.~\ref{fig:SChi} we show the estimated diffusion coefficient $D_{\ell_*}(\gamma)$ as a function of $\gamma$ across cutoff lengths $\ell_*$, together with the unitary value of \cite{PZ09}. To extract these diffusion coefficients we fit $\Delta S$ and $N$ to Eq.~\eqref{eq:JvsN} with fixed scaling exponent $\chi = 1$. $D_{\ell_*}(\gamma)$ appears linear in $\gamma$ for $\gamma \lesssim 0.05$. Linear extrapolation to $\gamma = 0$ puts the unitary diffusion coefficient $D(\gamma = 0)$ somewhat above the value of \cite{PZ09}; we again attribute this to the imprecision of our small-$\gamma$ simulations.

\subsection{Integrable XXZ Model}

Next, we investigate spin transport of the integrable XXZ model with anisotropy $0.0 \leq \Delta \leq 2.0$. We fix the magnitude of the artificial dissipation at $\gamma / \tau = 10.0$ for all cut-off lengths $\ell_\ast \in \{ 1, 2, \ldots, 5\}$. This artificial dissipation reduces the expectation value of operators longer than $\ell_*$ almost to zero, so it only allows processes involving operators with length $\ell > \ell_*$ on timescales shorter than $\tau = 1$.

For each $\ell_\ast$ and $\Delta$, we plot the the scaled average spin current $J / \Delta S$ as a function of the system size in Fig's \ref{fig:CJvsNweak} (anisotropic case $\Delta \ne 1$) and \ref{fig:CJvsNiso} (isotropic case $\Delta = 1$). We then display the scaling exponents $\chi$ as a function of anisotropy $\Delta$ across cutoff lengths $\ell_*$ in Fig.~\ref{fig:Cchi}.

\subsubsection{Generic anisotropic case $(\Delta \ne 0,1.0)$}\label{sss:generic-anisotropic}

\begin{figure}[t]
\includegraphics[width=\linewidth]{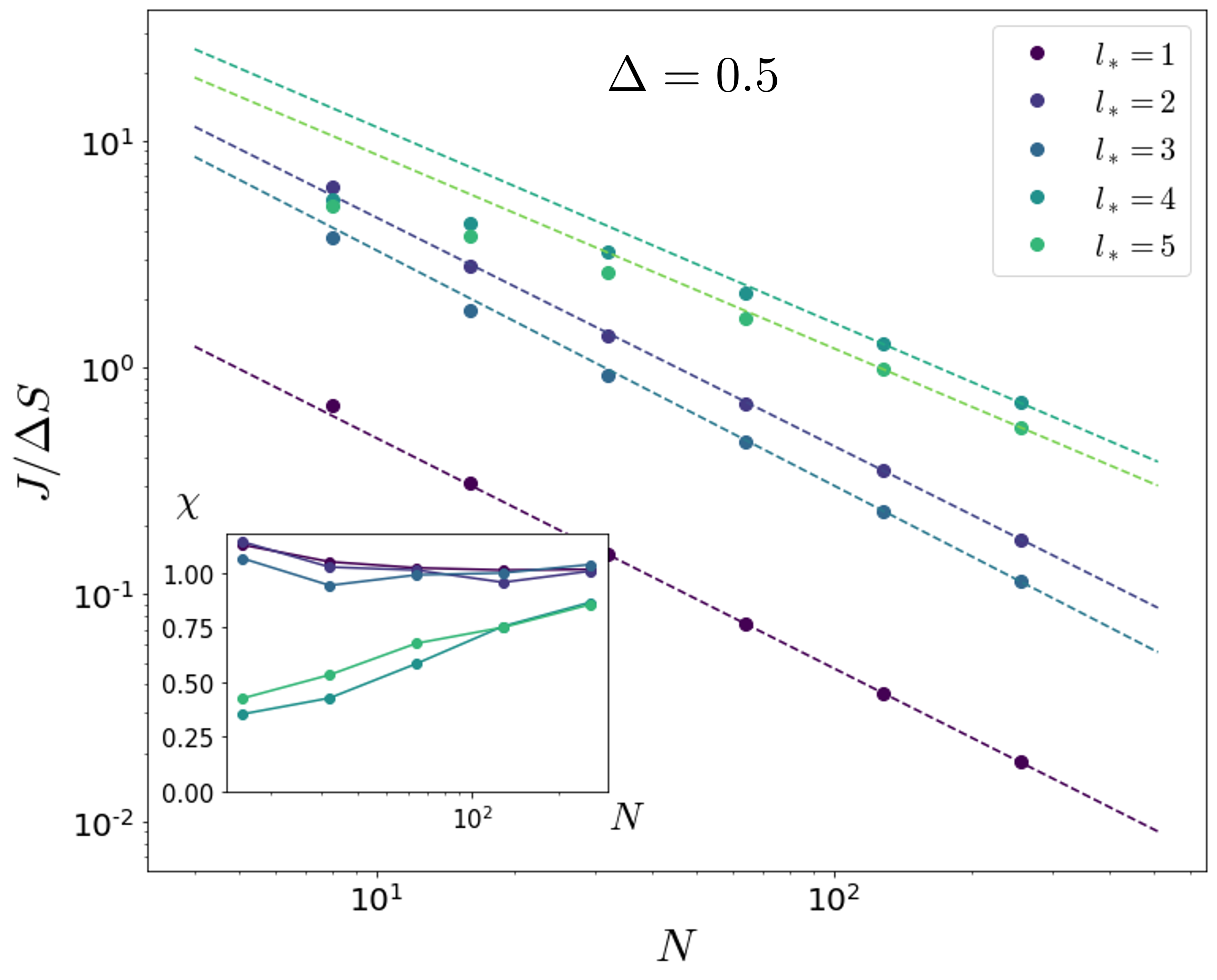}
\includegraphics[width=\linewidth]{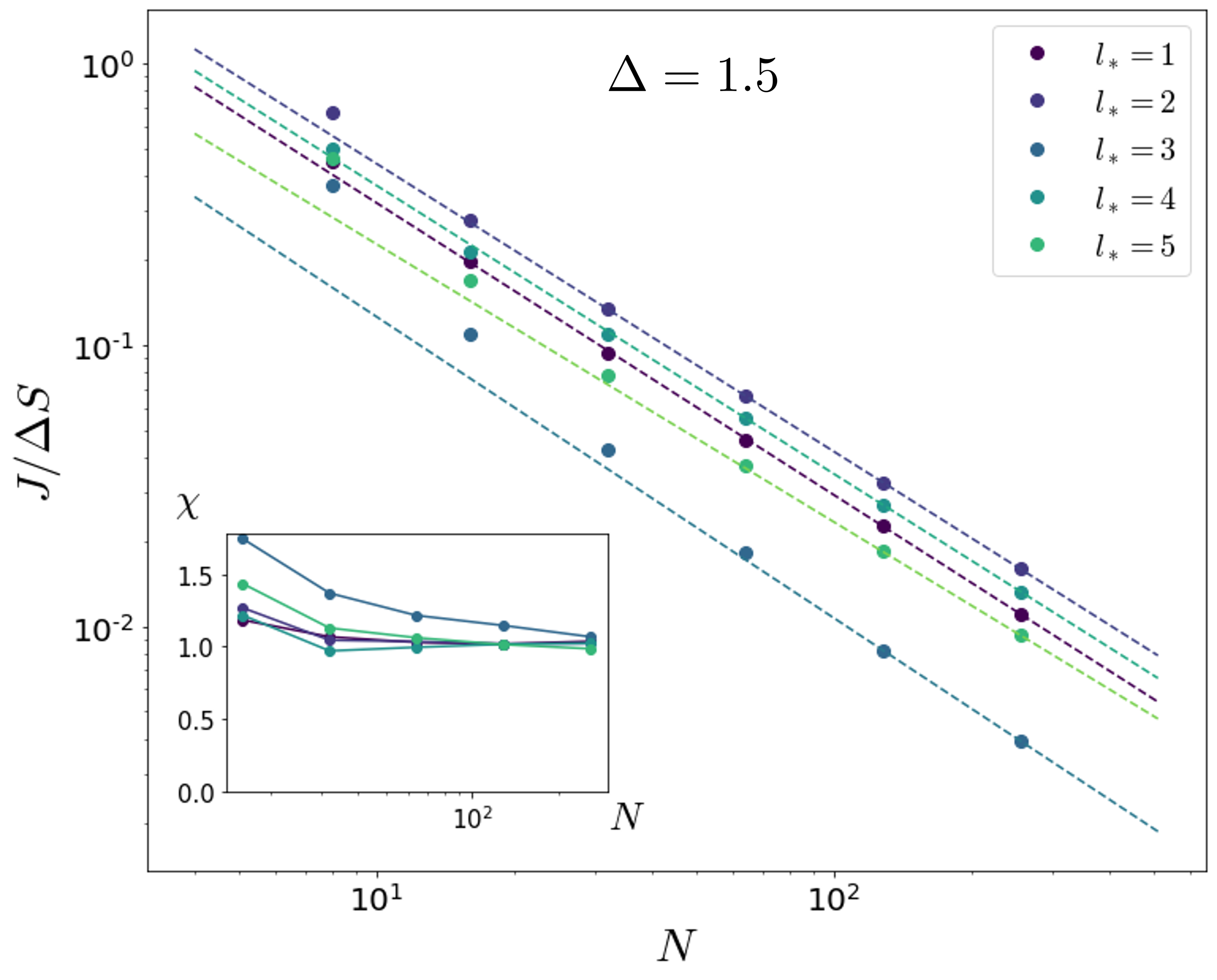}
\centering
\caption{Scaled average spin current of the clean Heisenberg XXZ model as a function of system size $N$ at $\Delta = 0.5$ (\textbf{top}) and $\Delta = 1.5$ (\textbf{bottom}). The dashed lines are best power law fittings corresponding to the (modified) NESS expectation values using data for $N\geq 100$.}
\label{fig:CJvsNweak}
\end{figure}

\begin{figure}[t]
\includegraphics[width=\linewidth]{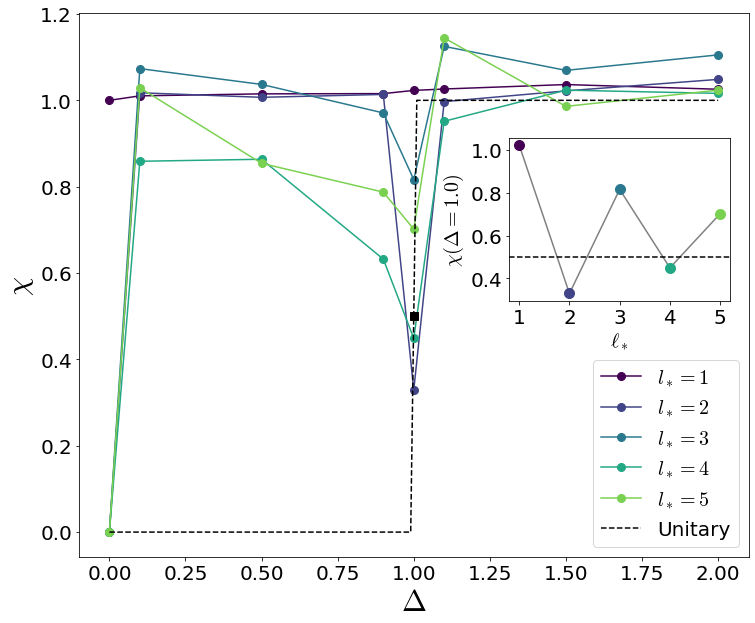}
\centering
\caption{Scaling exponents $\chi$ of the modified NESS from the anisotropic XXZ model without the disorder. The black dashed line represents the value without DAOE~\cite{Zni11, LZP17}.}
\label{fig:Cchi}
\end{figure}

Fig. \ref{fig:CJvsNweak} shows how the scaled average spin current depends on system size $N$ and cutoff length $\ell_*$ for weak anisotropy $\Delta = 0.5$ (top) and strong anisotropy $\Delta = 1.5$ (bottom).

For weak anisotropy, the unitary system displays ballistic transport. When strongly perturbed with DAOE it displays two kinds of behavior: for $\ell_* \le 3$ the system is nearly diffusive at all length scales, while for $\ell_* \ge 4$ it displays transient superdiffusive transport, with apparent exponent $\chi < 1$ at short length scales, before approaching diffusion at long length scales. We speculate that for $\ell_* \ge 4$ even strong artificial dissipation causes only weak scattering between quasiparticles.

For strong anisotropy the unitary model is diffusive. When strongly perturbed with DAOE it retains that diffusive behavior in the long-system limit. At short length scales, though, it displays apparent subdiffusive behavior with apparent exponent $\chi > 1$. This transient behavior is not monotonic in cutoff length.

Fig.~\ref{fig:Cchi} shows the scaling exponent $\chi$ as a function of the anisotropy $\Delta$. For $\ell_* = 1$ the modified NESS is diffusive (scaling exponent $\chi = 1$) at every anisotropy $\Delta$. But for $\ell_* > 1$ the modified NESS's behavior is much richer. It is ballistic at the free fermion point $\Delta = 0$, just like the unitary NESS, but even small interactions cause DAOE to push the system to diffusion. (Recall that in Fig.~\ref{fig:CJvsNweak} we saw the scaling exponent $\chi$ approaching the diffusive $\chi = 1$ as system size $N$ increased.) At the isotropic point, $\Delta = 1$ DAOE preserves some of the superdiffusive behavior of the unitary model. We discuss both $\Delta=0$ and $\Delta=1$ further in the next sections.

In this section we have worked at large $\gamma = 10$. We briefly discuss small $\gamma$ in App.~\ref{app:D_XXZ}.

\subsubsection{Free fermion case ($\Delta=0$)}

\begin{figure}[t]
\includegraphics[width=\linewidth]{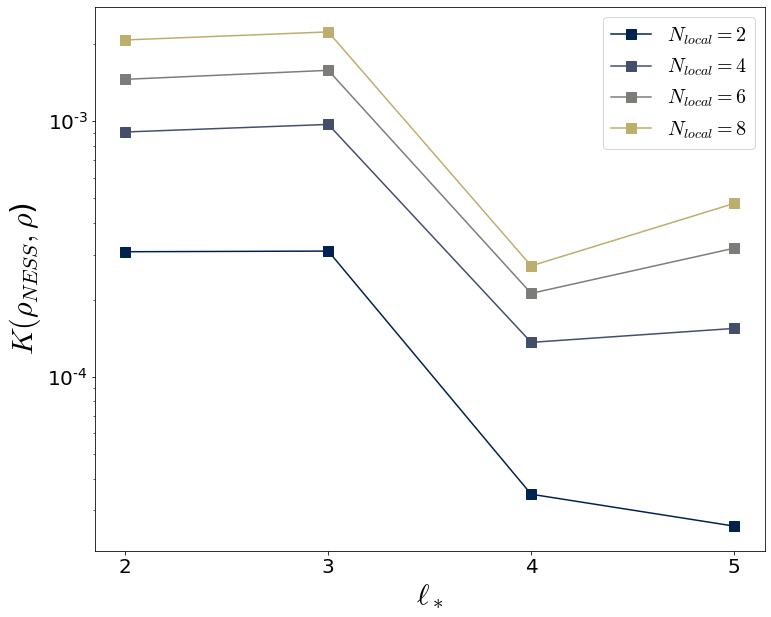}
\centering
\caption{Trace distance $K(\rho_{NESS}, \rho)$ between trial density matrices and the local NESSs obtained as reduced density matrices of the full NESS of the free fermion case for various local system sizes $N_{local}$. The local NESSs are extracted from the full NESS of $N = 48$ and $\chi = 32$.}
\label{fig:LocalNESS}
\end{figure}

As can be seen from left-most point in Fig.~\ref{fig:Cchi}, the $\Delta=0$ case retains ballistic transport provided $\ell_* >1$. This has a simple explanation in terms of the allowed NESS in the model. Most of the free fermion conserved quantities are quite complex in the spin language (see App.~\ref{app:jw}) But there is a special exception: the local current $J_i$, when summed over all lattice sites, is conserved. Moreover, because this operator is a sum of 2-site operators, it is also exactly preserved by DAOE. So a linearized steady state of the form $I + \sum_i a_i J_i$ is an exact NESS since it is preserved by DAOE and commutes with $H$. Provided $\ell_* >1$, we thus have a family of NESS that reproduce the ballistic transport of the unitary limit. This presumably explains the ballistic value seen in Fig.~\ref{fig:Cchi}.

To test this presumption, we study the structure of the NESS in more detail. In reality, we expect the true unitary NESS to have a form like
\begin{equation}\label{eq:max-ent-ness}
\rho_{NESS} \propto e^{- \sum_i a_i J_i} + \cdots
\end{equation}
with the $a_i$ uniform. This is the maximum-entropy state at zero energy and an appropriate current. where $\cdots$ denote corrections from other operators that are negligible for our purposes. The $a_i$ are typically small, so the linearized form in the previous paragraph is a good local approximation, but there are non-negligible corrections to the linearized form when the system size is larger. Fig.~\ref{fig:LocalNESS} shows a comparison of the trace distance between the postulated NESS of Eq.~\eqref{eq:max-ent-ness} and the NESS obtained from DAOE. Here, the trace distance between two density matrices $\rho_A$ and $\rho_B$ is defined as
\begin{equation}\label{tr_dist}
    K(\rho_A, \rho_B) = \frac{1}{2} \tr(\sqrt{(\rho_A - \rho_B)^2}).
\end{equation}
We see that the agreement is quite good already for $\ell_* = 2,3$ and further improves for $\ell_*=4,5$. This particular pattern of improvement arises because the expansion of $\rho_{NESS}$ only has operators of even weight, so when $\ell_*>1$ we preserve all the weight two operators but not weight four operators, and when $\ell_*>3$ we preserved all the weight two and four operators.

To summarize, when $\Delta=0$ we showed that DAOE preserves a family of NESS exhibiting ballistic transport whenever $\ell_* >1$. We further showed in Fig.~\ref{fig:LocalNESS} that these are the steady states realized in the actual converged simulation. Hence, we see again that the fact that DAOE preserves the summed current operator as a symmetry is crucial to recovering the unitary physics.

\subsubsection{Isotropic case $(\Delta = 1.0)$}

\begin{figure}[t]
\includegraphics[width=\linewidth]{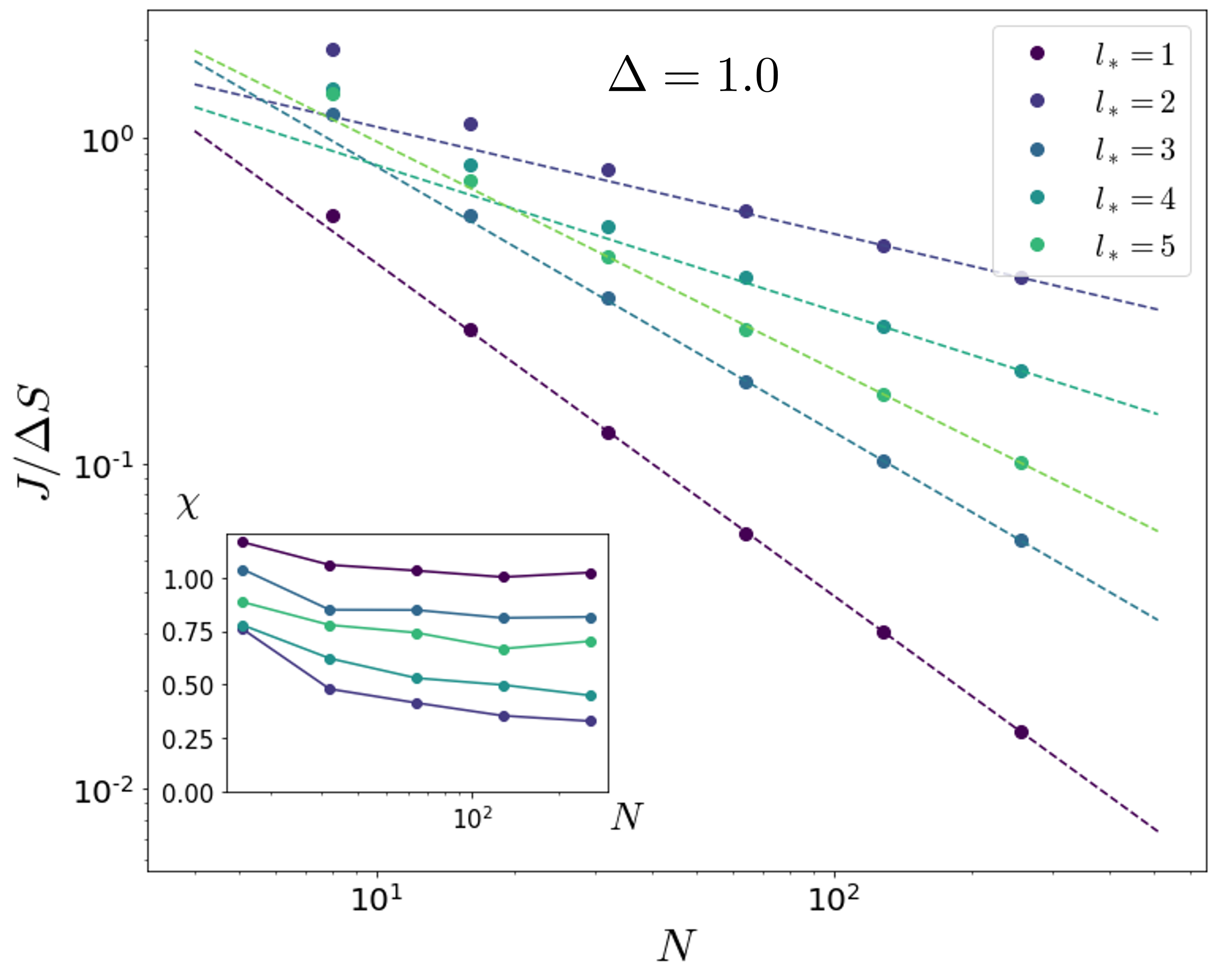}
\centering
\caption{Scaled average spin current of the clean Heisenberg XXZ model as a function of system size $N$ at $\Delta = 1.0$. The dashed lines are best power law fittings to corresponding the (modified) NESS expectation values.}
\label{fig:CJvsNiso}
\end{figure}

At the isotropic ($\Delta = 1.0$) point the model has an onsite $SU(2)$ symmetry. The unitary model is superdiffusive with exponent $\chi = 0.5$~\cite{LZP17, LZP19, DM20}. SU(2)-symmetric Hamiltonian perturbations do not appear to break superdiffusion down to diffusion \cite{DGV+21}, although this unexpected stability is believed to be a finite-size effect. The appropriate effective field theory is diffusive, and classical $SU(2)$-symmetric models generically display diffusion \cite{GDC+21}---though some classical models show finite-length transport faster than the asymptotic diffusion. $SU(2)$-symmetric dissipation, however, gives a system with at most logarithmic corrections to diffusion \cite{DGV+21}.

A priori, then, one expects DAOE to make the system diffusive. For $\ell_* = 1$ this is what we see (in both Fig.~\ref{fig:Cchi} and Fig.~\ref{fig:CJvsNiso}): the system has diffusive transport scaling exponent $\chi \approx 1$ for lengths $N \ge 32$. But for $\ell_* \ge 2$ the system appears to converge to superdiffusive transport powers $\chi < 1$. We believe this convergence is only apparent: that DAOE induces a quasiparticle scattering length longer than the system sizes $N \le 256$ we consider. The curious dependence of the exponent on $\ell_*$ may be a numerical signature of the contribution of different quasiparticle types to transport.

\subsection{Disordered XY Model}

\begin{figure*}[t!]
\begin{minipage}{\textwidth}
\includegraphics[width=\linewidth]{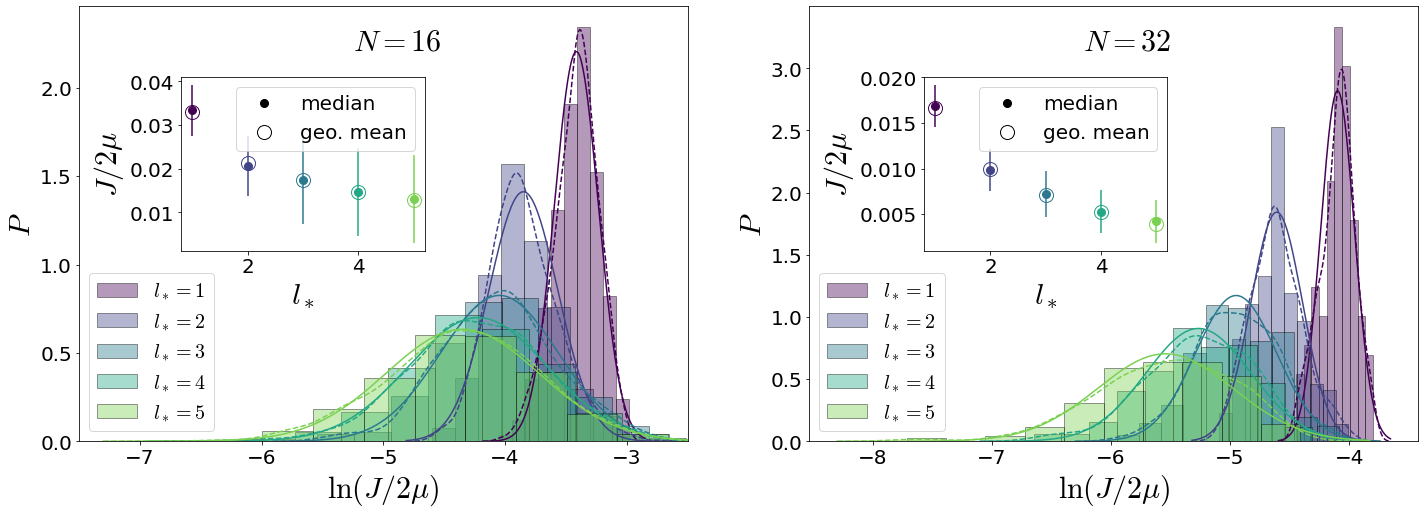}
\end{minipage}
\centering
\caption{Probability distribution of logarithm of the scaled spin current $\ln{(J/2\mu)}$ of the modified NESS for system size $N = 16$ and $N = 32$. The disorder strength is $h = 2.0$ for both system sizes. Each solid and dashed line represents a continuous probability curve and the best normal distribution fitting for the given parameter, respectively.}
\label{fig:JALdist}
\end{figure*}

Finally, we explore spin transport in the disordered XY model under strong operator weight dissipation. We consider a Hamiltonian
\begin{align}\label{eq:XY}
    H = \sum_j \left( \sigma^x_j \sigma^x_{j+1} + \sigma^y_j \sigma^y_{j+1} \right) + \sum_j h_j \sigma^z_j
\end{align}
where the fields $h_j$ are chosen uniformly at random $[-h,h]$. We consider $0.5 \le h \le 3.5$; for larger disorders, NESS convergence times are too large to be tractable. We take the operator weight dissipation large: $\gamma = 10.0$ at timestep $\tau = 1.0$.

The disordered XY model maps to a disordered free-fermion model (cf App.~\ref{app:jw}), so it is an Anderson insulator. The system's dynamics are determined by its Anderson orbitals (localized single-particle eigenstates). The Anderson orbitals have characteristic width
\begin{align}
    \xi \sim
    \begin{cases}
    24 / h^2 & h \lesssim 2.0\\
    1/\ln h & h \gtrsim 2.0\;.
    \end{cases}
\end{align}
To the extent that an isolated Anderson insulator transports charge, it does so coherently: charge tunnels into an Anderson orbital on the left end and out from the same orbital on the right end. Because orbitals are localized, these tunneling rates are small; additionally, the Anderson orbitals' onsite amplitudes are log-normally distributed, so the tunneling rates (hence conductivities) will likewise be log-normally distributed.

But suppose sites in the middle of the Anderson insulator are connected to a bath. The bath can cause incoherent transitions between Anderson orbitals because the dissipation superoperator will have matrix elements between Anderson orbital density matrices $\ketbra{\varepsilon_j}{\varepsilon_j}$, $\ketbra{\varepsilon_k}{\varepsilon_k}$ for nearby orbitals $\ket{\varepsilon_j}, \ket{\varepsilon_k}$. These matrix elements vary. But the resulting distribution of local resistivities has finite moments, so the system is diffusive---not subdiffusive
\footnote{
    Other behavior is possible. If a fraction $p < 1$ of sites are connected to a bath, for example, the system will have runs of sites with no dephasing; runs of a given length will be exponentially rare but have exponentially large resistance, giving subdiffusion \cite{TS21}.
    }.

The DAOE projection operator $\mathcal D_{l_*, γ}$ of Eq.~\eqref{DAO} likewise causes incoherent transitions between Anderson orbitals, because it likewise has matrix elements between Anderson orbital density matrices $\ketbra{\varepsilon_j}{\varepsilon_j}$, $\ketbra{\varepsilon_k}{\varepsilon_k}$. But (at least for the large disorder) these matrix elements are only nontrivial for orbitals centered at sites $j,k$ separated by at least $l_*$, and they will go as the amplitude of an orbital at some site a distance $l_*$ away from the orbital's center site. Charge therefore tunnels a distance $l_*$ at rate $\Psi_{l_*}$ where $\Psi_{l_*}$ is a log-normally distributed random variable with mean
\begin{align}\label{eq:and-daoe-rho-mu}
    \mu = e^{-l_*/\xi}
\end{align}
and some variance $\sigma$ determined by $l_*$ and the disorder properties. This gives a local diffusion coefficient $D \sim l_*^2 \Psi_{l_*}$ and a resistivity
\begin{align}\label{eq:and-daoe-rho}
    \rho \sim D^{-1} \sim (l_*^2\Psi_{l_*})^{-1}\;.
\end{align}

\begin{figure*}[t]
\centering
\includegraphics[width=\linewidth]{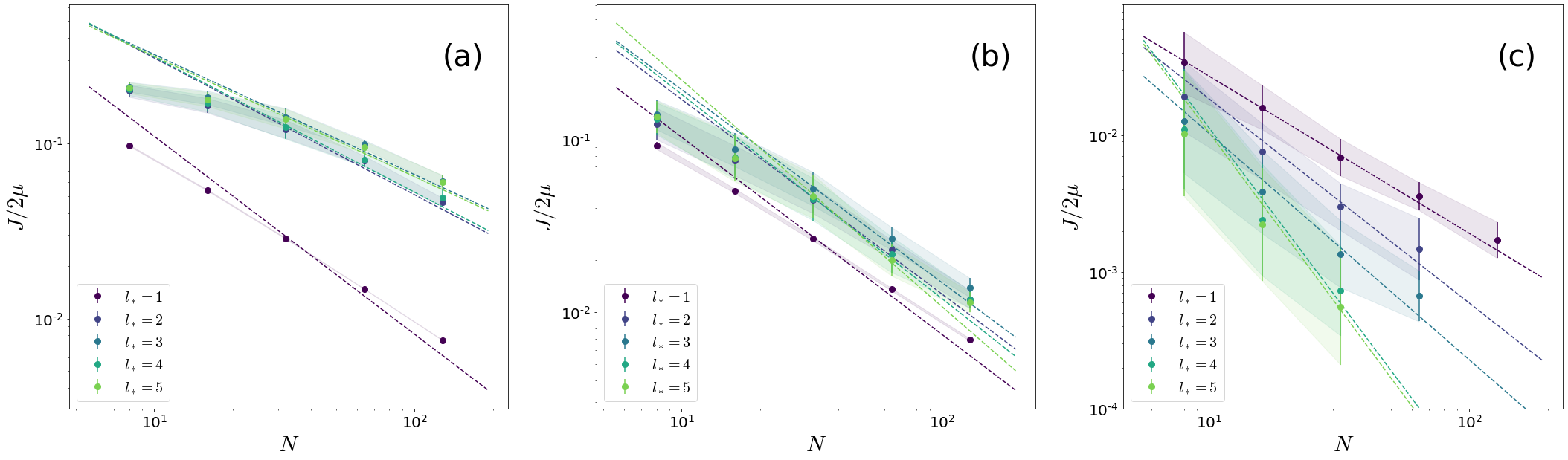}
\caption{The geometric mean and the associated error bar of spin current of the disordered XY model as a function of system size $N$. Each panel shows the result of different disorder strengths ((a) $h = 0.5$, (b) $h = 1.0$, (c) $h = 3.0$). The dashed lines are the best power law fittings corresponding to geometric means from the last three data points.}
\label{fig:ALJvsN}
\end{figure*}

We can therefore think of such a system---an Anderson insulator evolved under DAOE---as a network of resistors
of length $l_*$ and log-normal random resistance $R_j = l_*\rho$ distributed per \eqref{eq:and-daoe-rho} and \eqref{eq:and-daoe-rho-mu}. The total resistance is
\begin{align}\label{eq:R-total}
    R = \sum_{n = 1}^{L/l_*} R_n\;.
\end{align}
If the system is long enough, it will behave diffusively. The log-normal distribution has finite moments, so the resistance will be
\begin{align}
    \langle R \rangle \propto L/l_*\;,
\end{align}
with the constant given by the mean of the log-normal distribution, and the realization-to-realization variation will shrink as $1 / \sqrt{L/l_*}$. But for short systems, the sum \eqref{eq:R-total} is dominated by the largest individual resistance, which scales as
\begin{align}
   R_{\max} \sim e^{\mu + \sqrt{c L}}
\end{align}
for the mean $\mu$ of \eqref{eq:and-daoe-rho-mu} and some $l_*$ and $h$-dependent $c$
\footnote{
    This comes from standard results on extreme values of the normal distribution; see \cite{HF06} example 1.1.7.
    }.
Consequently, an ensemble of systems may look subdiffusive, even across more than a decade of length, even though the systems become diffusive in the large-size limit.

In Fig.~\ref{fig:JALdist} we show the distribution of log end-to-end steady-state currents for two different system lengths, $N = 16$ and $N = 32$. In each case, a kernel density estimator (solid line) matches a best-fit Gaussian (dashed line) well, and the median and geometric mean agree closely. In each case the median current decreases as $l_*$ increases, but for $l_* \gtrsim 3$ the decrease is more noticeable in the $N = 32$ distribution than the $N = 16$ distribution. This indicates that at $N = 16$ at least we are seeing substantial finite size effects.

In Fig.~\ref{fig:ALJvsN} we show the geometric mean currents
\begin{align}
    J_{\mathrm{geom}} = \exp{\langle \ln J/2\mu \rangle}
\end{align}
as a function of system size, across $l_*$ and disorder width $h$. In the limit of large system size, the distribution of total resistances will show central-limiting behavior, and approach a Gaussian with variance increasing slower than the mean. But at the system sizes we treat the distribution is still broad (e.g. in Fig.~\eqref{fig:JALdist} the $N = 32, l_* = 5$ current distribution spans about a decade). The arithmetic mean is therefore less enlightening than the geometric mean. The geometric mean currents appear to show power law scaling in system size---but this is because our data only span slightly more than a decade.

\begin{figure}[t]
\includegraphics[width=\linewidth]{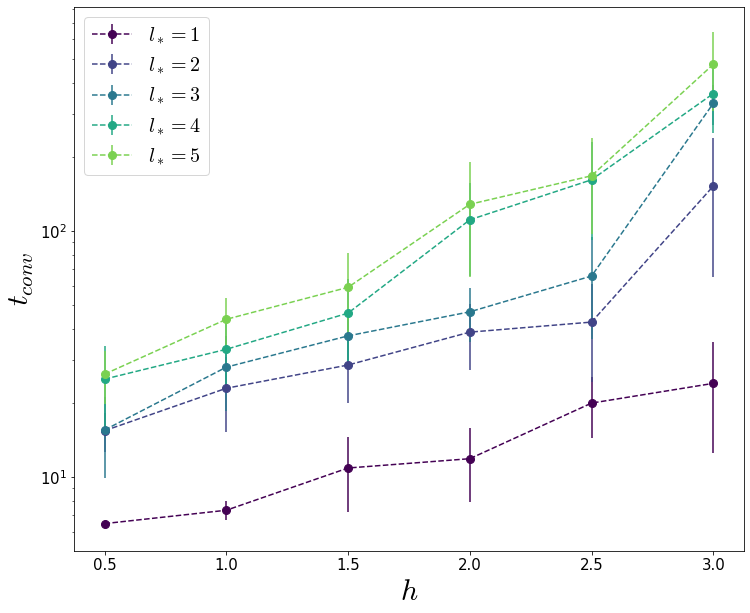}
\centering
\caption{Convergence time for various operator cut-off lengths $\ell_\ast$ and the disorder strengths $h$ for fixed system size $N = 32$.}
\label{fig:A5}
\end{figure}

Fig.~\ref{fig:A5} shows convergence times as a function of $h$ for system size $N = 32$. We consider the simulation converged in time when
\begin{align}
    \frac 1 J \frac{dJ}{dt} < 10^{-4}
\end{align}
where time is in units of coupling, and the site-to-site variation in current is less than the average across sites. We see a broadly exponential increase in convergence times with $h$.

\section{Discussion}

We have used non-equilibrium steady states of boundary-driven Lindblad systems to study how operator weight dissipation changes the scaling behavior of transport. We used as our test cases members of the XXZ family of models in three flavors: chaotic, integrable, and Anderson localized. We found that operator weight dissipation pushes the system towards diffusive transport scaling---except to the extent that the dissipation preserves some crucial symmetry or conserved quantity of the underlying dynamics. In the clean XXZ case at the isotropic point ($\Delta=1$), simulations with DAOE display anomalous transport similar to the underlying integrable dynamics, and at the free-fermion integrable point ($\Delta = 0$) it displays ballistic transport for $l_* \ge 2$ because it preserves the underlying model's single-particle momentum conservation and resulting Drude peak. In the disordered XY model, DAOE simulations resemble the localized behavior of the underlying system to the extent that DAOE preserves the system's Anderson orbitals.

Our central result is that even quite strong operator-weight dissipation does not necessarily change the scaling behavior of a system's transport, provided it preserves the system's symmetries. From an effective field theory point of view this is a gratifying result. The premise of the effective field theory approach to hydrodynamics is that a system's dynamics are characterized up to some $O(1)$ numbers by its symmetries and related conservation laws; our work tests precisely this contention. Moreover, since recent numerical methods for hydrodynamics (including DAOE, but also a number of other methods) also take this as their premise, our work offers supporting evidence for that approach.

But our results for the isotropic point of the XXZ model remain a mystery from this effective field theory point of view. From that point of view one would expect the operator-weight dissipation to break integrability and consequently turn the system into a generic system characterized simply by its symmetries, in this case $SU(2)$; such a system would be diffusive. We found that the superdiffusion at the isotropic point appeared robust to operator-weight dissipation, though this may have been a finite-size effect.

One interesting direction is to extend the method to fermionic systems where the length and/or weight of a fermionic operator string can be defined in a similar way to the spin case. As an application, it would be interesting to revisit free-fermion integrable or near-integrable models with technology, without the extra complications of a Jordan-Wigner transformation. This sort of operator weight dissipation might be applied to non-local models like the SYK model as well, although tensor network methods are not currently available in that case (but see \cite{SS22} for recent progress in the context of sparse models). 

It is also interesting to compare and contrast DAOE with more physical models of dissipation. For example, for various one-dimensional spin models, strong enough dephasing results in normal diffusion for various original transport types including ballistic transport to localization~\cite{Zni10, MAC+13, ZH13, ZMC+16, TS21}. We find that the effect of operator weight dissipation on open system dynamics is closely tied to the symmetries it preserves.

\section*{Acknowledgments}

CDW and YY thank the U.S. Department of Energy (DOE), Office of Science, Office of Advanced Scientific Computing Research (ASCR) Quantum Computing Application Teams program, for support under fieldwork proposal number ERKJ347.

\bibliographystyle{apsrev4-2}
\bibliography{DAOEbib}

\appendix

\begin{figure*}
\centering
\includegraphics[width=\linewidth]{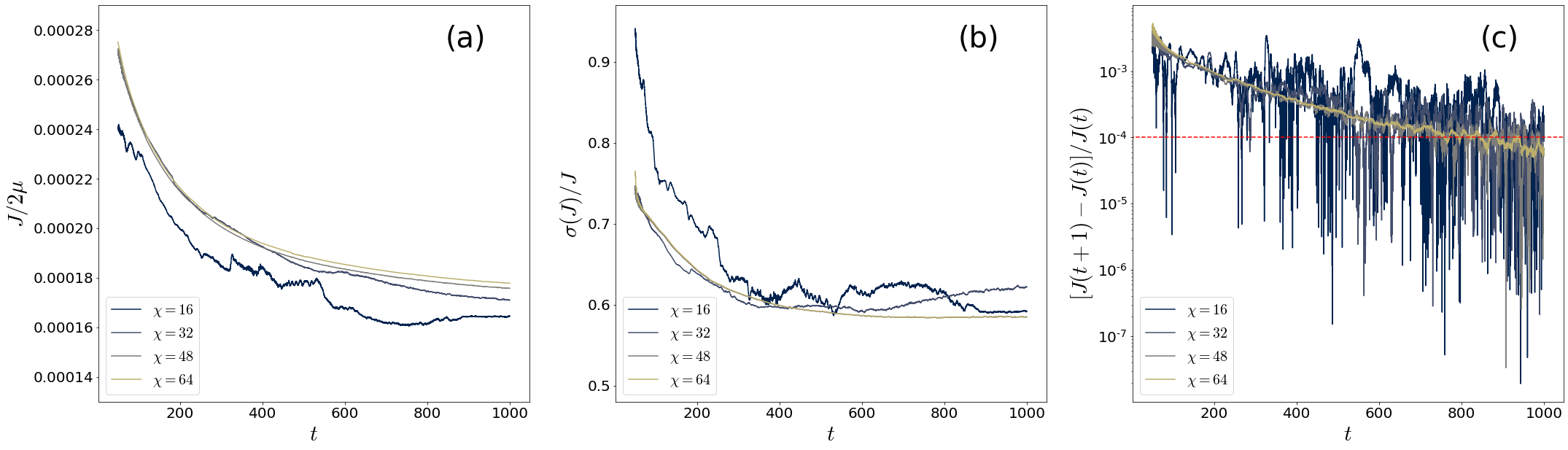}
\caption{Convergence in time to an approximate NESS of the disordered XY model with parameters $N = 32$, $h = 3.0$ and $\ell_\ast = 5$. (a) Current divided by bias as a function of time showing four different bond dimensions. We see approximate convergence in bond dimension. (b) Site-by-site variation of the current normalized by the average current as a function of time. (c) Change in current after one coupling time as a function of time. Our convergence criterion is that this normalized change is less than $10^{-4}$. This is achieved after approximately $t=800$.}
\label{fig:A3}
\end{figure*}

\section{Conserved Quantities in the XY Model}\label{app:jw}

Here we review the well-known construction of conserved quantities in the XY model obtained via Jordan-Wigner transformation to a non-interacting fermion problem.

In an XY chain of finite length, one can define fermion creation and annihilation operators as
\begin{equation}
    c_r^\dagger = \prod_{r' < r} \sigma^z_{r'} \frac{\sigma^x_r + i \sigma^y_r}{2}
\end{equation}
and
\begin{equation}
    c_r = \prod_{r' < r} \sigma^z_{r'} \frac{\sigma^x_r - i \sigma^y_r}{2}.
\end{equation}
For $r\neq r'$, it follows directly that
\begin{equation}
    \{c_r , c_{r'}^\dagger\} = 0,
\end{equation}
and for $r=r'$, we have
\begin{equation}
    \{ c_r , c_r^\dagger \} = \frac{1}{4} \{ \sigma^x_r - i \sigma^y_r, \sigma^x_r + i \sigma^y_r \} = 1.
\end{equation}

The Hamiltonian,
\begin{equation}
    H = - J \sum_r \left( \sigma^x_r \sigma^x_{r+1} + \sigma^y_r \sigma^y_{r+1} \right),
\end{equation}
in the fermion representation becomes
\begin{equation}
    H = - J \sum_r \left( c_{r+1}^\dagger c_{r} + c_r^\dagger c_{r+1} \right).
\end{equation}
From this representation, it is clear that we have an extensive set of conserved quantities given by 
\begin{equation}
    n_k = c_k^\dagger c_k
\end{equation}
where
\begin{equation}
    c_k = \sum_{r} \frac{e^{i k r}}{\sqrt{N}} c_r.
\end{equation}

It is instructive to convert these conserved quantities back into the spin language. First, we write them in terms of the position basis creation/annihilation operators,
\begin{equation}
    n_k = \sum_{r,r'} \frac{ e^{-ik(r-r')} }{N} c_r^\dagger c_{r'}.
\end{equation}
Next, we need the following identity for $c^\dagger_r c_{r'}$ valid for $r>r'$,
\begin{equation}
    c_r^\dagger c_{r'} = \frac{\sigma^x_r + i \sigma^y_{r}}{2} \left[ \prod_{r'' = r'+1}^{r-1} \sigma^z_{r''}\right] \frac{\sigma^x_{r'} + i \sigma^y_{r'}}{2}.
\end{equation}
Hence, a pair of fermion operators separated by $\ell$ sites gets mapped to a diameter (and weight) $\ell+2$ spin operator. The conserved quantities are in turn superpositions of all possible pairs of fermion operators. 

In the spin language, all non-trivial conserved quantities besides the charge and energy are superpositions that include many high weight operators. Therefore, any DAOE-like scheme will necessarily badly damage all such non-trivial conserved quantities. The non-interacting character of the model is strongly modified, so it is not surprising that the results tend toward a more generic diffusive behavior.

This analysis must be modified, however, in the presence of quenched disorder. This is because such disorder tends to localize the fermions, and the conserved quantities, which are local in momentum space at zero disorder, are expected to evolve with increasing disorder towards operators which are more local in position space. Indeed, in the extreme limit of very strong disorder, it is just the fermion number on every site that is conserved. In this case, DAOE will not disrupt such local conserved quantities and we can expect the physics of localization to be better captured than the physics at weak disorder. This is essentially what is observed in Fig.~\ref{fig:JALdist}.

\section{NESS Convergence}\label{app:conv}

\subsection{NESS Convergence for Different Initial States}

\begin{figure}[h]
\includegraphics[width=\linewidth]{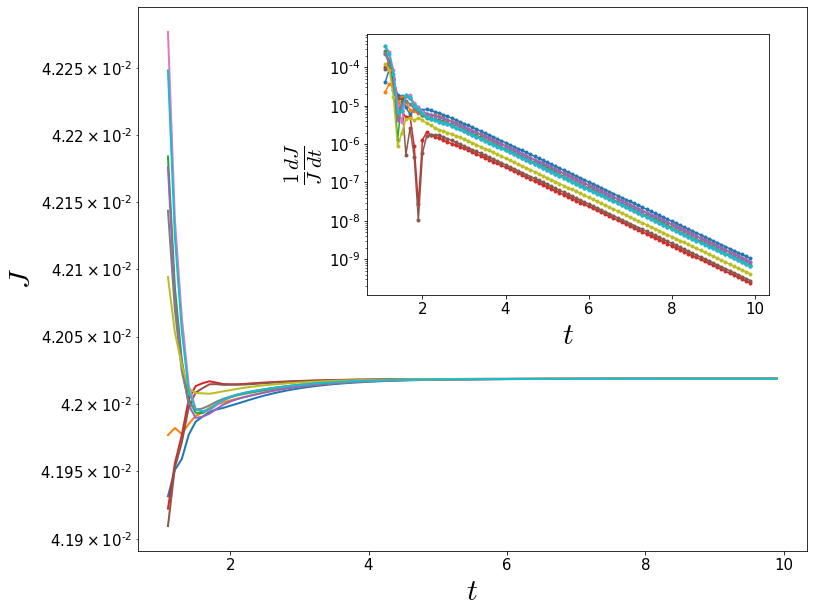}
\centering
\caption{Numerically exact calculations for the XXZ model with $\Delta = 0.5$ and system size of $N = 9$. Each colored line corresponds to a random product initial state.}
\label{fig:UniqueNESS}
\end{figure}

In the main text, we assumed that the systems in our study are under the appropriate conditions of the GKLS equation to find the corresponding NESS. Although we have not determined explicit forms of operators in Ref.~\cite{Nig19}, we tested the uniqueness of NESS with numerical experiments. Here, we prepared different random product density operators as initial states for the XXZ model with $\Delta = 0.5$. The numerically exact calculation for a small system size suggests that those random states converge to the same NESS (Fig.~\ref{fig:UniqueNESS}). We believe that similar behaviors are expected for the general XXZ model with different hamiltonian parameters and larger systems with truncations resulting in approximated NESSs.

\subsection{Clean XXZ Model}

In this section, we present the convergence of the modified NESS with bond dimension and trotter time step for non-disordered models in our study.  In our simulations, we consider that the convergence with the time is achieved if the relative error of spin current in one characteristic time $[J(t - 1) - J(t)] / J(t)$ is below $10^{-4}$.

First, the convergence with Trotter step size is represented in Fig. ~\ref{fig:A4} (a). In the main text, we used $\delta t = 0.1$ for all numerical calculations. We find that $\delta t = 0.1$ results show only $2\%$ difference compared to the smallest $\delta t = 0.025$ cases. It suggests that our choice of the Trotter time step correctly describes the given model's physics, given that a smaller Trotter step normally helps the simulation have a better approximation.

Next, the convergence with the bond dimension is shown in Fig. ~\ref{fig:A4} (b) According to our general observation, achieving the NESS convergence becomes more difficult as the anisotropy parameter $\Delta$ gets stronger. It is notable that the hardest case among our simulation parameter ($N = 256$, $\Delta = 2.0$ and $\ell_\ast = 5$) with $\chi = 64$ case also converges very well, demonstrating that the relative error is less than $2\%$ in comparison with the next largest bond dimension $\chi = 48$. The result implies that DAOE-assisted NESS converges with a relatively smaller bond dimension for the clean XXZ model.

\subsection{Disordered XY Model}

In the main text, we encountered apparent subdiffusive transport in the disordered XY model. In this slow dynamical regime, accessing an accurate NESS with a reliable convergence is difficult to achieve due to both expensive space and time computational complexities. Here, we describe the convergence of the modified NESS with respect to several simulation parameters.

Since a finite time evolution always approximates NESS in a practical calculation, accomplishing a tolerable error is the most important aspect of the simulation. Fig. ~\ref{fig:A3} (a) illustrates time evolution of scaled spin current of the modified NESS up to $t = 1000$. Similar to the non-disordered case, we set the same convergence criteria we employed for the clean XXZ model.

A similar trend is observed for the DAOE-NESS combined approach in accordance with the theory of DAOE, which states that artificial dissipation significantly reduces the required bond dimension to express a quantum state in question. One can also confirm that bond dimension $\chi = 64$ shows good convergence for the simulation with the longest cut-off length $\ell_\ast = 5$. Meanwhile, the shortest cut-off length $\ell_\ast = 1$ simulation only requires $\chi = 16$ to find a NESS convergence to a similar level (data not shown). Because the entanglement growth is more suppressed when applying DAOE with a shorter cut-off length, the quantum state can be efficiently represented with a smaller bond dimension. Meanwhile, a larger disorder and a longer cut-off length results in a shorter convergence time (Fig. ~\ref{fig:A5}) for fixed system size, as expected. Roughly, the convergence time exponentially growing function of both $\ell_\ast$ and $h$. This trend severely limits the accessible parameter regime of the study. However, we could not find any clear relation between bond dimension and convergence time.

Typically, site-to-site fluctuation of local spin current remains even though the average value converges. This site-to-site fluctuation also tends to slightly decrease by employing a larger bond dimension, however, it hugely depends on the disorder strength. This is measured by the scaled standard deviation $\sigma / J$ (Fig. ~\ref{fig:A3} (b)). The number of samples $M$ for each case is therefore determined by the statistical uncertainty $\sigma / J \sqrt{M}$, which is around 0.1 for the most difficult cases. For example, the $h=3.0$ case has $\sigma / J \sim 0.6$, and it follows that the required $M$ is around 25 to meet our criteria.

\begin{figure*}
\centering
\includegraphics[width=\linewidth]{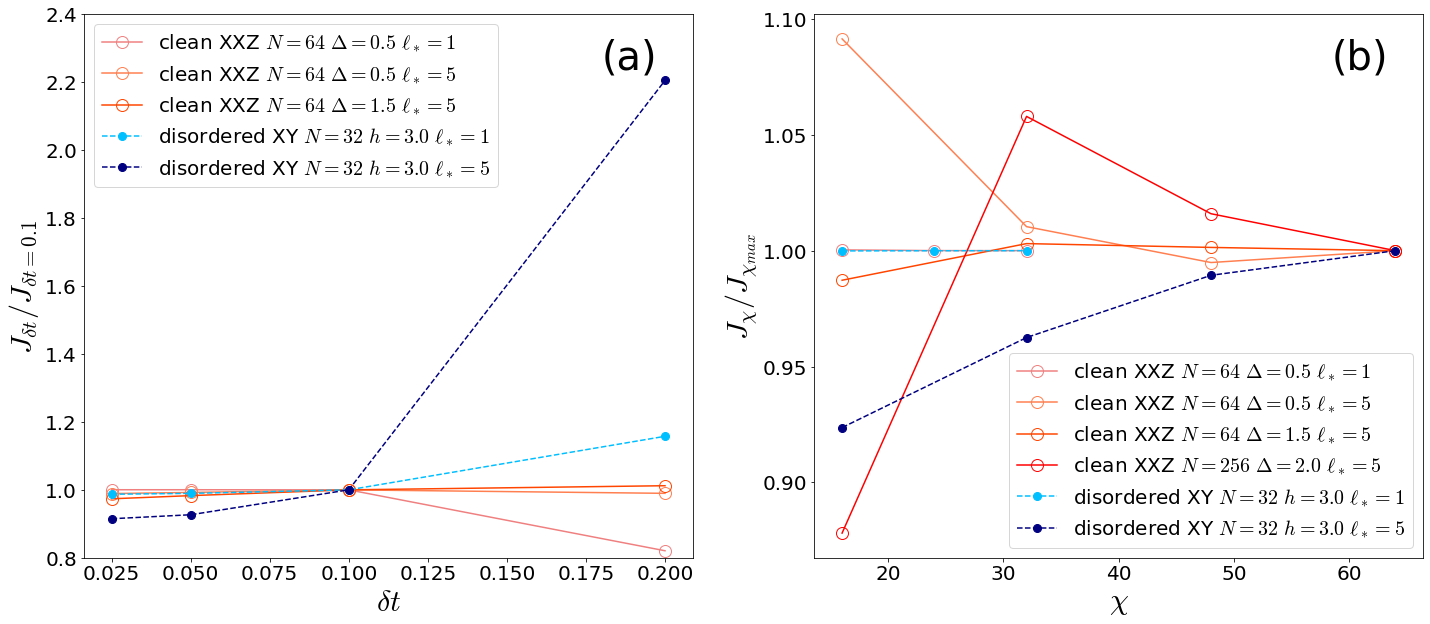}
\centering
\caption{(a) NESS convergence with Trotter time step. All data is extracted with bond dimension $\chi = 32$ for $\ell_\ast = 1$ and $\chi = 64$ for $\ell_\ast = 5$ .(b) NESS convergence with bond dimension. Here, the Trotter step size is fixed at $\delta t = 0.1$.}
\label{fig:A4}
\end{figure*}

It is clear that a smaller Trotter step $\delta t$ results in a more accurate NESS approximation, but choosing an appropriate $\delta t$ is necessary to reduce the total time complexity. In the main text, we used the second-order Suzuki-Trotter decomposition in the TEBD algorithm with $\delta t = 0.1$. The Hamiltonian contains at most 2-site operators, this even-odd type decomposed time evolution is expected not to change the physical properties of the model. The characteristic NESS convergences with the trotter step are shown in Fig. ~\ref{fig:A4} (a), for both shortest ($\ell_\ast = 1$) and longest ($\ell_\ast = 5$) operator cut-off lengths. The spin current seems to converge as we use smaller $\delta t$. The relative difference for the smallest $\delta t = 0.025$ case is about $1\%$ and $8\%$ for $\ell_\ast = 1$ and $\ell_\ast = 5$ respectively, compared to our usual value $\delta t = 0.1$ for the main text.

\subsection{Weakly Dissipated NESS}

In the main text, we frequently treat transport of the modified NESS under strong operator weight dissipations where subsequent dynamics is mostly governed by short operators. Analogous to the original study in Ref. ~\cite{RVP20} and diffusive case in the main text, it is deserving to explore the behavior of the modified NESS near the unitary limit. In particular, a reliable NESS calculation for a system with strong disorder and slow dynamics is notoriously hard. To investigate the potential improvement of NESS quality in subdiffusive transport physics, we choose the XY model with the Fibonacci disorder, which is realized by the deterministic disorder. In this model, the Fibonacci sequence gives the disorder $h$ or $-h$ for each site (See ~\cite{VZ19, CPP+21} for a detailed construction). Previous studies suggest that the model has varying anomalous transport; the scaling exponent $\chi$ monotonically increases as the disorder strength gets stronger. Here, we apply time-dependent $\gamma \sim 1/t$, trying to recover the unitary dynamics at a late time, otherwise, the setup is the same as in the main text.

\begin{figure}[h]
\includegraphics[width=\linewidth]{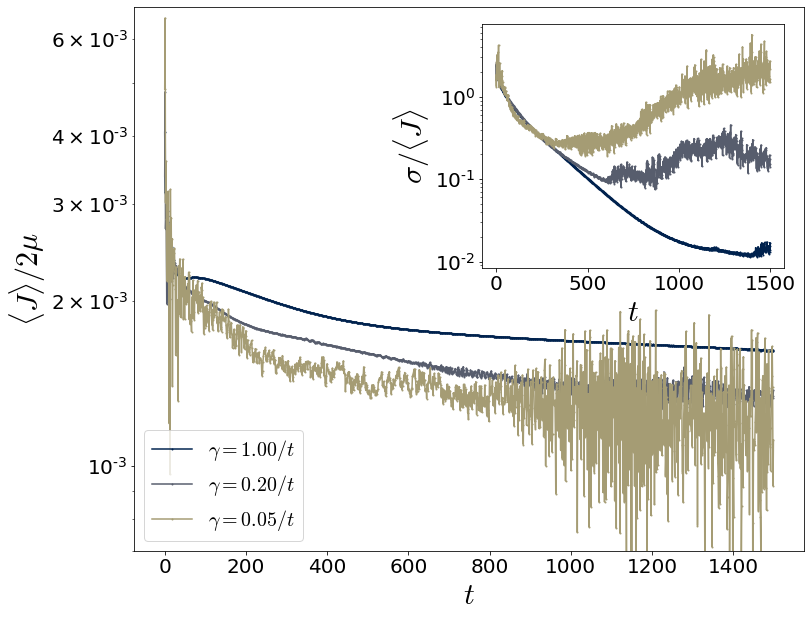}
\centering
\caption{Convergence of the scaled average spin current $\langle J \rangle / 2 \mu$ obtained from the modified NESS with weak dissipations for the XY model with the Fibonacci disorder ($N = 89$ and $h = 2.0$). The inset shows corresponding scaled standard deviation $\sigma / \langle J \rangle$ of $\langle J \rangle$.}
\label{fig:A1}
\end{figure}

Fig.~\ref{fig:A1} shows the time-evolving $\langle J(t) \rangle / 2 \mu$ for three dissipation strengths for the model with $h = 2.0$ whose (non-dissipative) transport is subdiffusive ($\chi \sim 1.17$). The expectation value is sensitive to the (stronger) early time dissipation before overcoming the entanglement barrier and slowly relaxing to the unitary dynamics. It appears that there remains the accumulative effect of the early time despite the rapid decrease in dissipation. Meanwhile, stronger dissipation results in smaller scaled standard deviation of the expectation value, $\sigma / \langle J \rangle$ (inset of Fig.~\ref{fig:A1}). The result shows a clear improvement of convergence time in the strongest $\gamma$ case as well. We have no clear relation of the tradeoff between the accuracy and the quality of convergence, however, the result offers a potential usage of this approach to approximate the unitary dynamics for such systems.

\section{Operator Weight Distribution for Small Systems}

\begin{figure}[h]
\includegraphics[width=\linewidth]{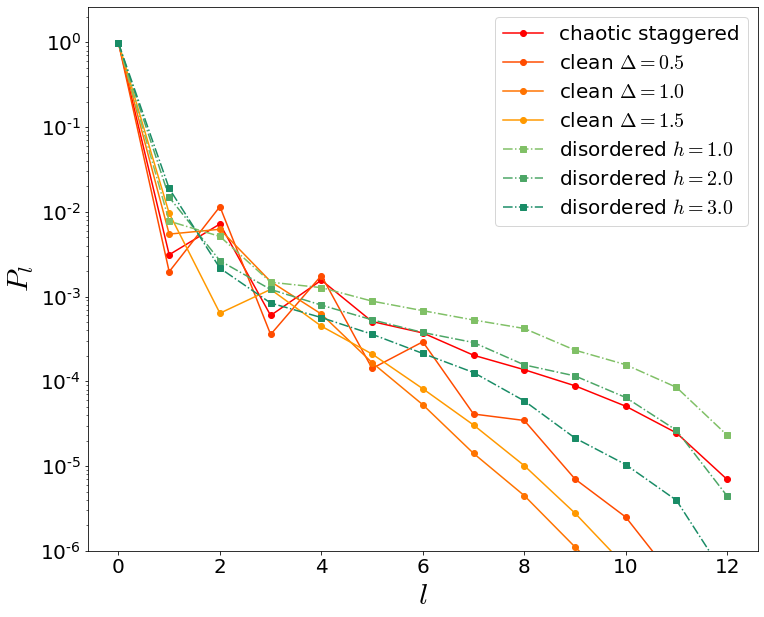}
\centering
\caption{Operator weight distribution for several parameters having different transport types as a function of length $l$.}
\label{fig:A2}
\end{figure}

The transport coefficients, $D$, and $\chi$, experience both non-monotonic and monotonic behavior in $\ell_\ast$ depending on the system in question. Among many possible factors, the operator weight distribution of the NESS can be directly related to those phenomena. For a small system size ($N=12$), it is possible to obtain the exact operator weight distribution of the NESS for several different parameters presented in the main text (Fig.~\ref{fig:A2}). In general, the trivial operator has almost of weight and the distribution exponentially decreases in length. However, non-monotonic distribution is observed for the ergodic systems, especially for short operators, which might be responsible for similar behaviors in the transport coefficients. On the other hand, the distribution has an exponential decrease trend for the localized models. The decreasing trend strengthens as the disorder increases, suggesting that the reinforced subdiffusive transport in a short length scale is connected to the operator weight distribution and the localization length. This investigation supports our rough analysis of the resultant non-monotonic or monotonic behavior of the transport coefficients in $\ell_\ast$ of the (modified) NESS.

\section{Energy diffusion coefficients for the XXZ model at $\Delta = 1.5$}\label{app:D_XXZ}

\begin{figure}
\includegraphics[width=\linewidth]{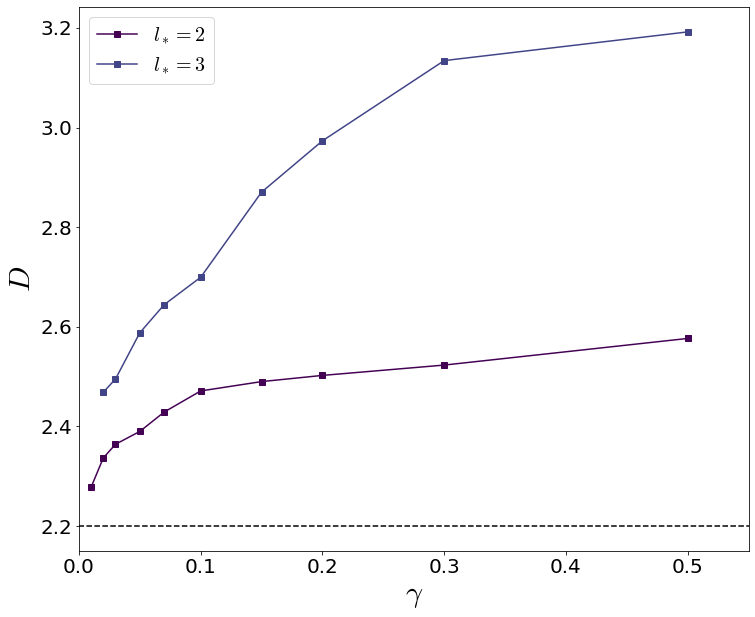}
\caption{Plot of diffusion constant $D$ of the anisotropic XXZ model with $\Delta = 1.5$ from various dissipation strengths $\gamma$. $D$ is extracted by using the best $1/L$ fitting curve, which is obtained from the system size $60 \leq L \leq 100$. Bond dimension is 64 and the convergence in the bond dimension is about 2\%. Other simulation parameters are similar to the main text. The black dashed line is the diffusion constant calculated by similar NESS setting without dissipation in \cite{Zni11}.}
\label{fig:D_XXZ}
\end{figure}

In Section \ref{sss:generic-anisotropic} we discussed how DAOE affects transport in the integrable anisotropic XXZ model at large artificial dissipation $\gamma = 10$. The XXZ model is diffusive for $\Delta > 1$; in this appendix we extract diffusion coefficients for the model at $\Delta = 1.5$.

Fig.~\ref{fig:D_XXZ} shows the diffusion coefficient as a function of $\gamma$ for $l_* = 2, 3$. Curiously, the diffusion coefficient is closer to the true value for $l_* = 2$ than for $l_* = 3$. We attribute this to the interplay between two effects. One the one hand, DAOE at any $l_*$ breaks the subtle integrability effects that give the true diffusion coefficient $D \approx 2.2$. Evidently, doing so increases the diffusion coefficient. On the other hand, the DAOE superoperator at $l_* = 2$ specifically decreases the energy current, because the energy current operator has nontrivial components consisting of three Pauli matrices, giving a correction to the general DAOE-caused increase of the diffusion coefficient. We expect that a better understanding would start with the effect of DAOE on the $\Delta = \infty$ kinetic picture of \cite{GV19, GHE13} (cf \cite{DGV+22}).

\end{document}